\documentclass[final,3p,times,twocolumn]{elsarticle}

\usepackage{amsmath,amssymb,amsfonts}
\usepackage{textcomp}
\usepackage{comment}
\usepackage{subcaption}
\usepackage{gensymb}
\usepackage{lineno}
\usepackage{adjustbox}
\usepackage{doi}

\journal{Ultrasonics}

\begin{document}

\begin{frontmatter}

\title{Large elements and advanced beamformers for increased field of view in 2-D ultrasound matrix arrays}

\author{Mick Gardner, Rita Miller, Michael L. Oelze}

\affiliation{organization={Electrical and Computer Engineering, University of Illinois at Urbana-Champaign},%Department and Organization
            addressline={306 N Wright St}, 
            city={Urbana}, 
            state={IL},
            postcode={61801},
            country={USA}}

\begin{abstract}
Three-dimensional (3D) ultrasound promises various medical applications for abdominal, obstetrics, and breast imaging. However, ultrasound matrix arrays have extremely high element counts limiting their field of view (FOV). Current reduced element count architectures, such as row-column arrays, diverging lenses, or sparse arrays, suffer from limited resolution and high side- and grating-lobe levels. This work seeks to demonstrate an increased field-of-view using a reduced element count array design. The approach is to increase the element size and use advanced beamformers to maintain image quality. The delay and sum (DAS), Null Subtraction Imaging (NSI), directional coherence factor (DCF), and Minimum Variance (MV) beamformers were compared. K-wave simulations of the 3D point-spread functions (PSF) of NSI, DCF, and MV display reduced side lobes and narrowed main lobes compared to DAS. Experiments were conducted using a multiplexed 1024-element matrix array on a Verasonics 256 system. Elements were electronically coupled to imitate a larger pitch and element size. Then, a virtual large aperture was created by using a positioning system to collect data in sections with the matrix array. Resolution and contrast was also assessed on a rabbit liver in vivo. Resolution was maintained using coupling numbers up to four, doubling the FOV while reducing the element count. The NSI and DCF beamformers demonstrated the best resolution performance in simulations, in a phantom with the virtual aperture, and in vivo on a rabbit liver. Our results demonstrate how larger matrix arrays could be constructed with larger elements, with resolution maintained by advanced beamformers. 
\end{abstract}


\end{frontmatter}

\section{Introduction}
Three- dimensional (3D) ultrasound is an attractive technique in the medical imaging community because it can provide full-volume views of a region of interest while being portable, safe, and capable of real-time imaging. Ultrasound volumetric imaging has been used in gall bladder volume estimation \cite{jouleh_comparison_2025}, active surveillance of thyroid cancer \cite{yan_inter-observer_2023}, breast tumor assessment \cite{shoma_ultrasound_2006}, and many other applications \cite{huang_review_2017,gilja_measurements_1999}. However, 3D ultrasound imaging usually requires a fully populated 2D matrix array. Such arrays have a prohibitively high number of array elements, e.g. 1024 elements in a 32x32 system. With an array pitch typically around one wavelength or less to prevent grating lobes during beam steering, the field of view (FOV) of such probes can be severely limited. Also, most ultrasound scanners are not equipped with enough input channels to receive data from such a high element count, so multiplexers have been used in research scanners to switch between sections of the array on different transmit/receive events \cite{yu_design_2020}. However, in this case, frame rates are limited by the need for multiple transmit/receive events to acquire all aperture data. With a 4x1 multiplexer, anywhere between 4 - 16 transmit events would be needed to acquire, for example, a single plane-wave angle \cite{chavignon_3d_2022}. 

To handle higher element counts, many clinical systems integrate so-called “microbeamformers” into the handles of the transducers \cite{savord_fully_2003,rothberg_ultrasound--chip_2021}. With microbeamforming, the delay-and-sum (DAS) operation is partially performed on ASICs inside the transducer handle on patches of elements. Then, only one wire per patch needs to go back to the scanner for the remaining time delay compensation. With this approach, probes have been designed with over 9,000 active elements \cite{acar_real-time_2014}. However, the micro-beamformers introduce deviations from ideal time delays due to static focusing and quantization \cite{zhao_error_2015}. Micro-beamformed patches often have a fixed focal depth to simplify the hardware, which is not always aligned with the dynamic receive focus when beamforming across patches \cite{zhao_error_2015}. Delay quantization is also introduced by the micro-beamformer architecture, which might use a sample-and-hold method that can only delay signals by discrete time intervals determined by an input clock \cite{yu_programmable_2010}. These quantizations can reduce the resolution and contrast of high-frame rate images, especially for larger micro-beamformed patches \cite{castrignano_impact_2025}.

Other approaches allow the FOV to extend beyond the footprint of the array. These approaches include convex and phased arrays \cite{kang_wide_2020}, as well as panoramas (also called extended field-of-view or EFOV) \cite{kim_extended_2003}. With convex and phased arrays, the field of view can extend beyond the probe footprint due to array curvature or beam steering, respectively. However, with this approach, maintaining resolution over the FOV is a challenge because scan lines become more spread out with depth. With panoramas, the field of view is extended by translating the probe to multiple locations, acquiring an image at each location, then applying some image registration method to align the different views \cite{poon_three-dimensional_2006, wachinger_three-dimensional_2007}. The issue with panoramas is they can only acquire static images, never real-time video data, because of the need to translate the probe to multiple locations. 

A few alternative approaches to array design allow for larger footprints, and thus larger FOVs, with a reduced element count. Extensive work has been done to explore sparse arrays as an option for reducing the element count of 2D ultrasound arrays \cite{ramalli_design_2022}. With sparsity, element spacing is allowed to increase far beyond the usual 1/2 - 1 wavelength in some kind of optimized pattern. The most serious drawback of sparse arrays is that they have reduced transmit power and SNR because of the low element count and small elements. 

Another example of a reduced element count design is a row-column addressed (RCA) array \cite{jensen_anatomic_2022}. These are fully populated matrix arrays where instead of addressing individual elements, entire rows or entire columns are accessed at once on a single wire. This effectively creates two orthogonal arrays of long, line elements, reducing the element count from N x N to N + N. To further increase the FOV with row-column arrays, diverging lenses have been placed over the entire aperture \cite{audoin_diverging_2022}, with specialized beamformers designed for such lensed arrays \cite{salari_beamformer_2025}. Researchers have also designed curved, toroidal RCA arrays to implement that divergence \cite{caudoux_curved_2024}. However, the main limitation of these arrays are that they cannot focus or steer along diagonals, they can only focus or steer in a cross (rows, then columns, or vice-versa). This means that resolution will be worse with RCA arrays compared to fully addressed arrays which can steer and focus along both directions at once, in both transmit and receive. 

Finally, large, square elements have been used in sparse arrays with ultrasound localization microscopy (ULM) \cite{favre_boosting_2022,favre_transcranial_2023,haidour_multi-lens_2025}. With very large elements, the directivity of elements raises the minimum F-number of the array, degrading resolution. In this case, researchers overcame this issue by placing a diverging lens over individual elements (as opposed to over the entire aperture) to widen the element directivity \cite{haidour_multi-lens_2025}. A wider directivity will introduce grating lobes into the point-spread function (PSF) of the imaging system when the array pitch is greater than one wavelength. The lens approach is successful with ULM because ULM displays the tracks of moving microbubbles over many thousands of frames, and the microbubble localization and tracking algorithms ignore grating lobes. However, for B-mode imaging, grating lobes create artifacts that can obscure other image details \cite{paul_side_1997,gardner_grating_2024}, meaning a lens which allows grating lobes to appear is not desirable.

The design we propose is a periodic matrix array with large, square elements. Rather than using a lens, we also propose the use of adaptive or non-linear beamformers for the task of regaining resolution lost by increased element directivity. Several beamformers have been studied for their effectiveness at improving resolution and contrast. The Null Subtraction Imaging (NSI) beamformer has been shown to greatly improve resolution in 2D B-mode images \cite{agarwal_improving_2019}, and reduce grating lobe artifacts \cite{kou_grating_2022,gardner_grating_2024}. The NSI beamformer was also recently implemented on a 2D array for 3D imaging \cite{yociss_null_2021}. Coherence factor beamformers have often been used for improving contrast, and they can have benefits to resolution as well. Recently, a Directional Coherence Factor (DCF) beamformer was proposed specifically for use in matrix arrays \cite{wu_directional_2025}, which calculates coherence on directional projections of matrix data, then combines directions for improved resolution. The Minimum Variance (MV) beamformer has also been shown to greatly improve resolution for 1D arrays \cite{synnevag_adaptive_2007}, but has not been extensively used on 2D matrix arrays due to computational complexity. Therefore, we implemented MV on the same directional projections as DCF to reduce the data size \cite{wu_directional_2025}. The resolution improvement makes these beamformers attractive for solving the grating lobe and directivity issues from a larger pitch and element size.

The goals of this paper are to demonstrate how larger elements can lead to larger apertures without increase to element count and to determine which beamformer is best for maintaining resolution with large elements. To test our approach of using large, square elements, we used a commercial matrix array and electronically coupled adjacent elements to act as if they were one element. To demonstrate a larger aperture, we also used a positioning system to acquire data from a virtual aperture where data was collected in quadrants. Images were beamformed with DAS, NSI, DCF, and MV beamformers for quality comparison.

\section{Background Theory}
This section will give a brief analysis on the effects of increased element size on the beam pattern of a 2D matrix array. The array beam pattern is the product of two factors: an array factor arising from element spacing, and an element factor (often referred to as the directivity) arising from element size. Mathematically, this can be expressed as 
\begin{equation}
    B(\theta) = H(\theta)G(\theta)
\end{equation}
where \(B(\theta)\) is the beam pattern, while \(H(\theta)\) and \(G(\theta)\) are the array factor and element directivity respectively. The far-field directivity of an element is given by its spatial Fourier transform. A square element can be represented by two orthogonal rectangle functions, as in 
\begin{equation}
    A(x,y) = rect\left(\frac{x}{W}\right) rect\left(\frac{y}{H}\right)
\end{equation}
where \(A(x,y)\) is the aperture function, \(W\) is the width and \(H\) is the height of the element. Because these two rectangle functions are independent, we can separate the 2D spatial Fourier transform of the aperture into two 1D spatial Fourier transforms along the x- and y- axes, leading to 
\begin{equation} \label{eq:fourier_independence}
\begin{aligned}
    \mathcal{F}\{A(x,y)\} &= \mathcal{F}_{x}\left\{rect\left(\frac{x}{W}\right)\right\} \times \mathcal{F}_{y}\left\{rect\left(\frac{y}{H}\right)\right\} \\
    &= Wsinc\left(\frac{Wk_x}{2}\right)\times H sinc\left(\frac{Hk_y}{2}\right)
\end{aligned}
\end{equation}
where \(\mathcal{F}\{\cdot\}\) denotes the spatial Fourier transform, and \(k_{x}\) and \(k_{y}\) are the wave numbers in the \(x\) and \(y\) directions. We note that Equation \ref{eq:fourier_independence} represents a one-way sensitivity, and we are assuming that transmit and receive sensitivities are identical due to acoustic reciprocity. From Equation (\ref{eq:fourier_independence}), increased \(W\) and \(H\) values create more narrow sinc functions, representing how larger elements are less sensitive to off-axis echoes. This directivity raises the achievable F-number of the array, which is given by [24]:
\begin{equation} \label{eq:minimum_fnumber}
    F^{\#} = \frac{z}{aperture} = \frac{1}{2\tan(\alpha)}
\end{equation}
where \(z\) is imaging depth and \(\alpha\) is the -3 dB angle of the directivity. This raised F-number will degrade the lateral resolution that can be achieved with the array when using DAS beamforming \cite{foiret_improving_2022}. It is desirable to choose the minimum F-number derived from the element directivity to improve resolution for a given element size. It has also been shown that choosing arbitrarily low F-numbers will introduce clutter and reduce contrast \cite{perrot_so_2021}. Therefore, all F-numbers chosen in this paper are derived from Equation \ref{eq:minimum_fnumber}, using the -3 dB point of the element directivity for \(\alpha\). Because the elements are square, we use the lateral width of the elements to calculate \(\alpha\) and assume symmetry in elevation. 

\begin{figure}
    \centering
    \includegraphics[width=\linewidth]{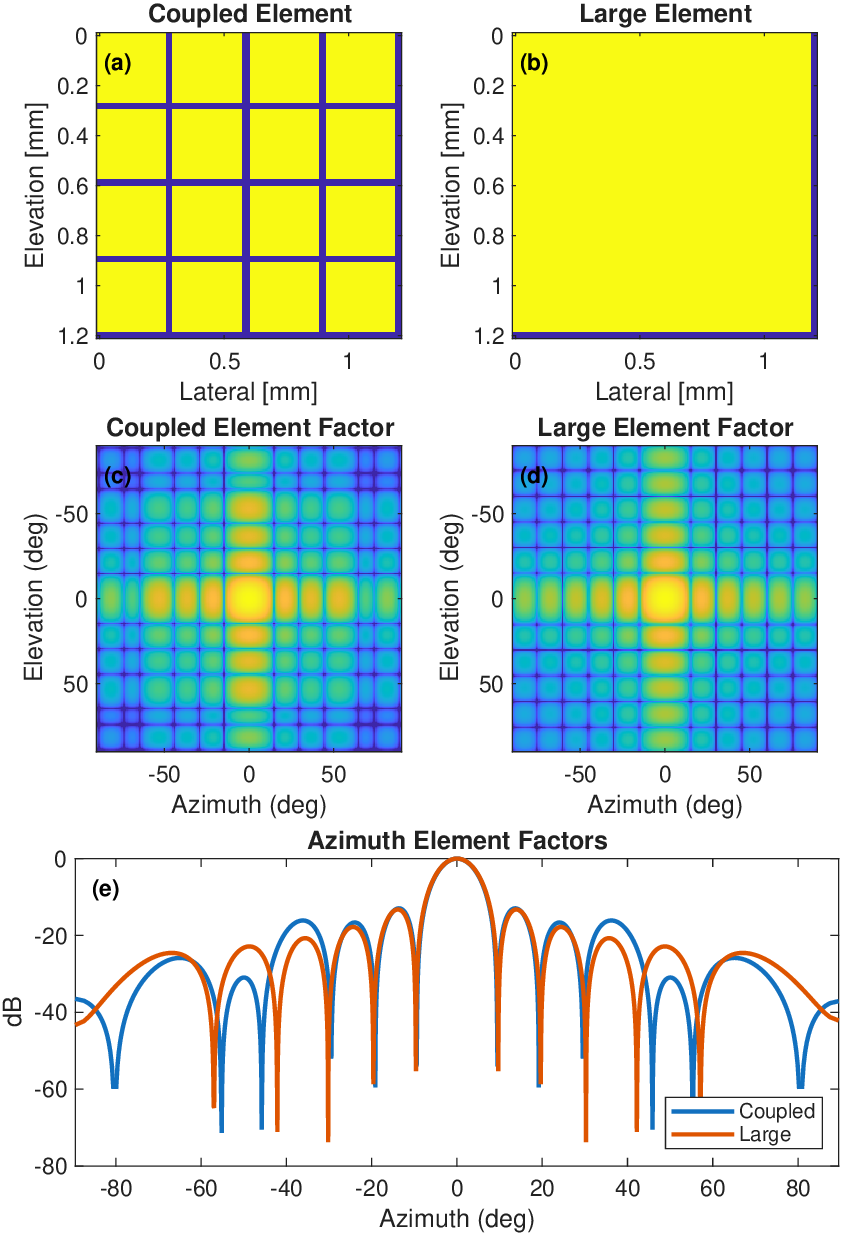}
    \caption{Representations of (a) a block of coupled elements and (b) and a large element, with corresponding Fourier transforms representing their directivities in (c) and (d). (e) is the cross section of the 2D Fourier transforms at an elevation of 0. Note there are only small differences in the side lobes, meaning coupled elements are a good approximation of the directivity of a large element, despite kerf gaps.}
    \label{fig1:coupled_element_factors}
\end{figure}

Lastly, for phantom and in vivo experiments, we are using coupled elements as an approximation of large elements. A block of coupled elements will have gaps, called the kerf, from the saw used to cut the piezo-electric elements. Figures \ref{fig1:coupled_element_factors}a-b display representations of a coupled element and a large element, while Figures \ref{fig1:coupled_element_factors}c-d display the corresponding 2D spatial Fourier transforms. Figure \ref{fig1:coupled_element_factors}e displays an azimuthal cross section of both directivities at an elevation angle of zero. Note that azimuth and elevation cross sections are identical due to symmetry. The spatial frequencies sampled by the 2D FFT were converted to arrival angles using the following equation, which can be derived from geometry:
\begin{equation}
    \theta = \sin^{-1}\left(\frac{kc}{2\pi f_{0} \Delta x}\right)
\end{equation}
where \(\theta\) is the angle of arrival, \(k\) is the spatial frequency sampled by the FFT, \(c\) is the sound speed (assumed to be 1540 m/s), \(f_{0}\) is the transmit frequency (set to 7.81 MHz), and \(\Delta x\) is the sample spacing (same for x- and y- axes). From these figures, it can be observed that the kerf gaps only create small differences in the side lobes, as well as nearly identical main lobes, suggesting that the directivity of a coupled element is a close approximation of a large element despite the kerf gaps.

\section{Methods}

\subsection{Simulations}
Simulations were conducted in K-wave \cite{martin_simulating_2016,treeby_rapid_2018} to examine the point-spread functions of the different beamformers on a large-element array. A total of four simulations were done, each with points at depths of 5 mm, 10 mm, 15 mm, and 20 mm. The four different simulations placed these points at different lateral/elevational locations: one for the center, one at 2.5 mm laterally, one at 2.5 mm elevationally, and one in the corner at 2.5 mm in both directions. Lateral and elevational symmetry are assumed for the simulation, so we use only one edge for each direction and one corner without loss of generality. In addition to multiple point locations, simulations were performed at different noise levels. White Gaussian noise was added to simulated channel data with amplitudes 0 (noiseless), 0.001, and 0.005. 

The simulated array had 64 elements in an 8 x 8 grid, with a pitch of 1.25 mm in both lateral and elevational directions and an element size of 1.20 mm in both directions. These dimensions match the size of coupling by four on the Vermon array used for phantom and in vivo experiments. The transmitted frequency was 7.81 MHz with a medium sound speed of 1540 m/s, making the wavelength 197 \(\mu\)m. The array parameters then translate to a pitch of 6.34 \(\lambda\), and an element size of 6.09 \(\lambda\), where \(\lambda\) is the wavelength. This element size yields a minimum F-number of 6.89 using Equation \ref{eq:minimum_fnumber}. 

\begin{figure*}
    \centering
    \includegraphics[width=\linewidth]{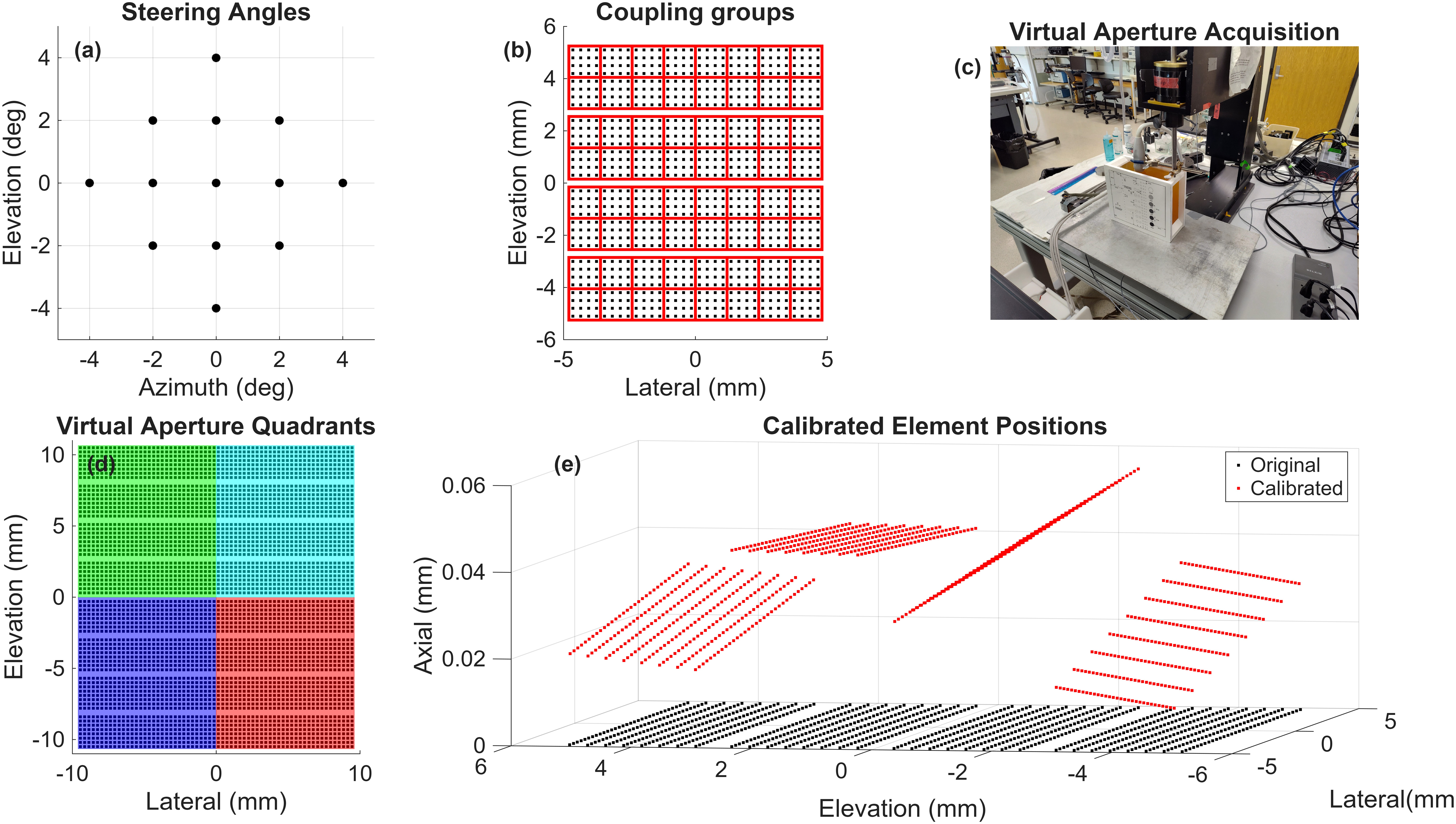}
    \caption{(a) Steering angles used in simulation, phantom, and in vivo experiments. Fewer diagonal angles were included due to more narrow directivity on the diagonals of square elements. (b) Example of coupling groups for a coupling number of 4. (c) Photograph of the Vermon array on the Daedal positioning system with the ATS phantom underneath. (d) Diagram of the four acquisition quadrants for the virtual aperture. (e) Calibrated element positions estimated using multi-lateration \cite{mccall_element_2024}.}
    \label{fig2:angles_coupling_vla_cal}
\end{figure*}

The plane wave transmission sequence consisted of 13 angles, arranged in a star pattern over azimuth and elevation as displayed in Figure \ref{fig2:angles_coupling_vla_cal}a. These angles were chosen based on the maximum steering range for the simulated element size. The 6.34 \(\lambda\) width led to a max steering angle of about \(4\degree\) in the lateral and elevational directions. Also, these elements were \(6.34\lambda\ \times \sqrt2=8.61\lambda\) along their diagonals, leading to a max steering angle of about \(3\degree\) diagonally. Therefore, we chose a star pattern of transmission angles, as opposed to a full square, so that fewer diagonal angles were included.

\subsection{Virtual Large Aperture}
A Vermon 1024-element 8 MHz matrix probe (Vermon S.A., Tours, France) was used to create the virtual aperture. The Vermon probe is made up of four panels of 8x32 elements each. Calibrated element positions were used in calculating transmit and receive time delays on the Vermon array. The calibration method was multi-lateration \cite{mccall_element_2024}. Briefly, this method involves taking hydrophone measurements to triangulate the position of each corner of individual panels. Once corner locations are estimated, the plane intersecting all corners is calculated and remaining elements are evenly spaced between corners in that plane. This assumes each panel is flat, but different panels may have different rotations or displacements relative to one another. The calibrated element positions are given in Figure \ref{fig2:angles_coupling_vla_cal}e. 

Element coupling on this probe was performed by grouping blocks of adjacent elements, averaging their transmission delays, then averaging their received channel data without applying a receive time delay. Within each block of coupled elements, the transmission delays were averaged so that all elements in that block fired at the same time. Then, received RF traces were averaged across elements without applying a time delay so that received signals superimpose. By coupling on transmit and receive, each block acted as if it was a single, large element. Table \ref{tab2:vla_counts_widths} lists coupling numbers with corresponding element counts, element widths, and minimum F-numbers (Eq. \ref{eq:minimum_fnumber}) for the virtual large aperture. Figure \ref{fig2:angles_coupling_vla_cal}b illustrates how elements would be grouped for a coupling number of 4. Because of the panel separation and multiplexing on the Vermon probe, coupling numbers were limited to 1 (i.e. no coupling), 2, and 4 to avoid coupling across panels. While a coupling number of 8 could have been attempted without coupling across panels, element sizes would have reached 12 \(\lambda\) with an acceptance angle of \(2\degree\). In that case, a significant portion of the field of view would only be visible to one element at a time, meaning no beamforming could be performed. Therefore, we omitted a coupling number of 8.

To create the virtual large aperture, the probe was attached to a positioning system, and programmed to move into four quadrants, as displayed in Figure \ref{fig2:angles_coupling_vla_cal}c-d. The same plane-wave angle set was used in phantom experiments as for simulations. The transmit delays were calculated for every element of the virtual aperture. Once the transmission profiles were calculated, the probe was set in one position, fired the transmission profile associated with the VLA elements in that position, and collected data. Then, the probe was moved to the next position, fired with the corresponding transmission delays, collected data, and so on until the full aperture data was acquired. Data was acquired in this manner on the wire and cyst targets in a CIRS Model 539 ATS General purpose phantom (Computerized Imaging Reference Systems, Norfolk, VA, USA). Two wire target datasets were acquired, one with the wires aligned laterally, and the other elevationally, to examine resolution in both dimensions. For FOV comparison, one dataset each for a wire and a cyst was collected with the Vermon array by itself as well. 

Before scanning the four quadrants, visual alignment was performed to make the probe as flat as possible when scanning. A plane-wave setup script was run on Verasonics to create a real-time B-mode display. The probe was kept a few millimeters above the boundary of the ATS phantom so the boundary was visible in the display. Then, the probe was rotated by hand in its clamp until the boundary appeared flat in both lateral and elevation directions. The probe was then lowered to be nearly touching the boundary. For the final alignment, a wire target was visualized in the C-scan, and the probe was rotated in Daedal until the wire appeared vertical. Each virtual aperture acquisition was repeated 3 times on separate occasions to include variability in the setup/alignment procedure in the results.

Finally, we opted to use the same angle set as the simulations (see Figure \ref{fig2:angles_coupling_vla_cal}a) for all element sizes because a coupling number of 4 on the Vermon array matches the element size of the simulation. Also, using the same angle set for all coupling numbers allowed a direct comparison of quality based only on element size and beamforming, even though a larger steering range was available with less coupling.

\subsection{Rabbit Liver}
Animal experiments were approved by the Institutional Animal Care and Use Committee (IACUC) at the University of Illinois Urbana-Champaign, protocol \#23062. One rabbit was anesthetizing using isoflurane. Then the hair on the abdomen was shaved and the rabbit's liver was scanned transabdominally with the Vermon array by itself. Due to the rabbit's breathing motion, in vivo acquisitions with the virtual aperture were not feasible. Three datasets were collected for coupling numbers of 1, 2, and 4 (i.e. nine total datasets), each containing 10 frames of plane-wave acquisitions. The same plane-wave angle set was used as previous experiments (see Figure \ref{fig2:angles_coupling_vla_cal}a). From these datasets, nine frames for each coupling number with visible blood vessels were chosen for quality metric estimation.

\subsection{Beamforming}
\subsubsection{Delay-and-Sum} \label{section:dasMethod}
Delay-and-sum (DAS) beamforming is the conventional method of beamforming which simply delays channel data based on transmit and receive distances and sums the delayed data \cite{perrot_so_2021}. Received radio-frequency (RF) data were demodulated into In-phase/Quadrature (IQ) data before beamforming. Beamforming was done on a pixel grid with spacing 200 \(\mu\)m in lateral, elevational, and axial directions. Once time delays were applied, transmissions were compounded coherently and elements were summed 
\begin{equation} \label{eq:beamformer_sum}
    y(\vec{r}) = \sum_{i=1}^{N} \sum_{k=1}^{M}  IQ_{i,k}\left(\tau(\vec{r},\vec{x_{i}})\right) e^{-j2\pi f_{0}\tau(\vec{r},\vec{x_{i}})}.
\end{equation}
In the above equation, \(y\left(\vec{r}\right)\) is the beamformed pixel value for pixel \(\vec{r}\), \(M\) and \(N\) are the number of transmissions and elements respectively, \(IQ_{i,k}\) is the delayed IQ data from element \(i\) and transmission \(k\), \(\tau\left(\vec{r},\vec{x_i}\right)\) is the time delay applied to the IQ data, and \(f_{0}\) is the center frequency of transmission.

Receive sub-apertures were determined using the minimum constant F-number approach described in \cite{perrot_so_2021}, where the -3 dB point of the element directivity determined the minimum F-number (Eq. \ref{eq:minimum_fnumber}). The lateral width of the elements was used to estimate the maximum acceptance angle.

\subsubsection{Null Subtraction Imaging}
Null subtraction imaging is implemented by beamforming with three different apodizations in parallel and incoherently summing the results \cite{agarwal_improving_2019}. These three apodizations are a zero-mean (ZM) apodization, and two direct current (DC) apodizations which are offset versions of the ZM. The ZM apodization creates a null at 0\(\degree\), broadside to the receive sub-aperture. The DC apodizations bridge the null but create similar side lobes in the beam pattern. Once beamforming is done with the three apodizations, the null is subtracted leaving a narrow main lobe and low side lobes \cite{agarwal_improving_2019}. The amount of offset, referred to as the DC offset, is a tunable parameter of NSI typically in the range 0.1-1. In our simulations and experiments, we set the DC offset to 0.5. Lower DC offsets result in narrower main lobes, lower side lobes, reduced grating lobes, and under-developed speckle \cite{agarwal_improving_2019,kou_grating_2022,gardner_grating_2024,kou_high-resolution_2024}.

To implement NSI on a 2D array, a directional approach was taken as described in \cite{yociss_null_2021}. In this approach, two zero-mean apodizations are made by taking an equal number of +1 and -1 along row and column directions of the receive sub-aperture. Once beamforming and envelope detection are done for each ZM apodization, the maximum value over each result is taken as a directional zero-mean, as in
\begin{equation} \label{eq:max_zm}
    E_{ZM} = \max\left(E_{ZM_{row}},E_{ZM_{col}}\right)
\end{equation}
where \(E_{X}\) represents the envelope of a beamformed signal. Beamforming is also performed with two DC offset apodizations for each ZM apodization. For a given ZM apodization, the DC offset apodizations are given by
\begin{equation}
    \begin{aligned}
        DC1_{X} &= ZM_{X} + dc\\
        DC2_{X} &= ZM_{X} - dc
    \end{aligned}
\end{equation}
where \(ZM\) represents the ZM apodization, \(dc\) represents the DC offset, \(DC1\) and \(DC2\) represent the DC apodizations, and \(X\) is a placeholder for direction. Once beamforming is performed for each DC offset apodization, the envelopes of the beamformed signals are averaged, as in
\begin{equation} \label{eq:avg_dc}
    E_{DC_{X}} = \frac{E_{DC1_{X}} + E_{DC2_{X}}}{2}.
\end{equation}
Then, a directional DC envelope is taken as the pixel-by-pixel maximum over both directions,
\begin{equation} \label{eq:max_dc}
    E_{DC} = \max\left(E_{DC_{row}},E_{DC_{col}}\right).
\end{equation}
Finally, the NSI image is obtained by subtracting the ZM envelope from the DC envelope,
\begin{equation} \label{eq:subtract_null}
    E_{NSI} = \left|E_{DC} - E_{ZM}\right|
\end{equation}
where the absolute value is taken to prevent negative envelope values.

\subsubsection{Directional Coherence Factor}
The Directional Coherence Factor (DCF) involves projecting matrix array data along azimuth and elevational dimensions, calculating a coherence factor for each direction, then multiplying those factors together \cite{wu_directional_2025}. Mathematically, the projected data can be written as
\begin{equation} \label{eq:projected_vectors}
    \begin{aligned}
        P_{i} &= \sum_{j=1}^{Q} S_{i,j} = \left[P_{1}^{AZ},...,P_{Q}^{AZ}\right]\\
        P_{j} &= \sum_{i=1}^{Q} S_{i,j} = \left[P_{1}^{EL},...,P_{Q}^{EL}\right]^{T}
    \end{aligned}
\end{equation}
where \(P_{i}\) is the projection onto azimuth, \(P_{j}\) is the projection onto elevation, \(Q = \sqrt{N}\) is the number of elements in azimuth and elevation, and \(S_{i,j}\) is time-delayed channel data from the element at row \(i\) and column \(j\). From these projected vectors, a coherence factor is calculated for each of them as 
\begin{equation}
    \begin{aligned}
        CF_{AZ} &= \frac{\left| \sum_{i=1}^{Q} P_{i} \right|^{2}}{Q \sum_{i=1}^{Q} \left| P_{i} \right|^{2}}\\
        CF_{EL} &= \frac{\left| \sum_{j=1}^{Q} P_{j} \right|^{2}}{Q \sum_{j=1}^{Q} \left| P_{j} \right|^{2}}.
    \end{aligned}
\end{equation}
Finally, the DCF is calculated by multiplying \(CF_{AZ}\) and \(CF_{EL}\):
\begin{equation}
    DCF = CF_{AZ} \times CF_{EL} = \frac{\left| \sum_{i=1}^{Q} \sum_{j=1}^{Q} P_{i} P_{j} \right|^{2}}{Q^{2} \sum_{i=1}^{Q} \sum_{j=1}^{Q} \left| P_{i}P_{j} \right|^{2}}.
\end{equation}
After the DCF value is calculated for each pixel, the pixel is multiplied by the DCF. We note that in the original method, the authors included diagonal projections \cite{wu_directional_2025}. However, for reduced computation, we opted to only include the row and column projections.

\subsubsection{Minimum Variance}
Inspired by the directional projections from DCF, we also performed Minimum Variance (MV) weighting on the projected vectors in Equation \ref{eq:projected_vectors} \cite{synnevag_adaptive_2007}. The optimal MV weights are given by 
\begin{equation} \label{eq:mv_weights}
    \textbf{w} = \frac{\hat{R}^{-1} \textbf{a}}{\textbf{a}^{H} \hat{R}^{-1} \textbf{a}}
\end{equation}
where \textbf{w} is the desired weighting vector, \(\hat{R}\) is an estimate of the spatial covariance matrix, and \textbf{a} is a steering vector set to all ones. The spatial covariance matrix was estimated on either direction using spatial smoothing as in 
\begin{equation} \label{eq:projected_covariance_matrices}
    \begin{aligned}
        R_{AZ} &= \frac{1}{Q-L+1} \sum_{l=0}^{Q-L} P_{i,l}P_{i,l}^{H}\\
        R_{EL} &= \frac{1}{Q-L+1} \sum_{l=0}^{Q-L} P_{j,l}P_{j,l}^{H}
    \end{aligned}
\end{equation}
where \(P_{i,l}\) and \(P_{j,l}\) were sub-vectors of the directional projections in Equation \ref{eq:projected_vectors}, defined as the following:
\begin{equation} \label{eq:spatially_smoothed_projections}
    \begin{aligned}
        P_{i,l} &= \left[ P_{l}^{AZ} , ... , P_{l+L-1}^{AZ} \right]\\
        P_{j,l} &= \left[ P_{l}^{EL} , ... , P_{l+L-1}^{EL} \right].
    \end{aligned}
\end{equation}
The sub-array size \(L\) was set to \(Q/2\) for both azimuth and elevation projections. In both cases, diagonal loading was also performed on the covariance matrix estimates:
\begin{equation}
    \hat{R} = R + \epsilon I
\end{equation}
where \(I\) is the identity matrix and \(\epsilon=\frac{1}{10L}\cdot trace\{R\}\) is the diagonal loading factor. After estimating the covariance matrices for either direction, weights were calculated with Equation \ref{eq:mv_weights}, resulting in a \(\textbf{w}_{AZ}\) and a \(\textbf{w}_{EL}\). Then, projected MV outputs were calculated as
\begin{equation}
    \begin{aligned}
        y_{AZ} &= \frac{1}{Q-L+1} \sum_{l=0}^{Q-L} \textbf{w}_{AZ}^{H} P_{i,l}\\
        y_{EL} &= \frac{1}{Q-L+1} \sum_{k=0}^{Q-L} \textbf{w}_{EL}^{H} P_{j,l}
    \end{aligned}
\end{equation}
and finally, directions were combined with
\begin{equation}
    y = \sqrt{y_{AZ}y_{EL}^{*}}
\end{equation}
where \(y\) is the final directional MV output, and \(^{*}\) denotes complex conjugate.

\subsubsection{Quality Metrics}
To assess the spatial resolution in simulation and phantom experiments, the full width at half-maximum (FWHM) of lateral and elevational profiles of point targets was estimated. The profiles were interpolated by a factor of 100 using spline interpolation for FWHM estimates. The rabbit liver scans lacked appropriate point/line targets on which to measure envelope FWHM. Therefore, the lateral and elevational width of the volume autocorrelation was used to estimate resolution. The autocorrelations of full volume envelopes were computed using the FFT. Then, FWHM estimates were made on the center lateral and elevational profiles of the autocorrelations. These profiles were also interpolated by a factor of 100 for FWHM estimates.

The contrast of volumes was estimated using the contrast ratio (CR), the contrast-to-noise ratio (CNR), and the generalized contrast-to-noise ratio (gCNR). These metrics are given by the following equations \cite{rodriguez-molares_generalized_2020}:
\begin{align}
    CR &= 10\log_{10}\left(\frac{\mu_{i}}{\mu{o}}\right)\\
    CNR &= \frac{\mu_{i} - \mu_{o}}{\sqrt{\sigma_{i}^{2} + \sigma_{o}^{2}}}\\
    gCNR &= 1 - \int_{-\infty}^{\infty} \min_{x}\left\{p_{i}(x),p_{o}(x)\right\}dx
\end{align}
In the above equations, \(\mu_{i}\) is the average envelope level inside some anechoic region of interest (ROI), and \(\mu_{o}\) is average envelope level of the background. Likewise, \(\sigma_{i}^{2}\) and \(\sigma_{o}^{2}\) are the variances of the envelopes in those regions. Finally, \(p_{i}\left(x\right)\) and \(p_{o}\left(x\right)\) are the probability distributions of the envelopes in the same regions. Histograms with 256 bins were used to estimate the distributions. In simulations, the main lobe to side lobe ratio (MLSLR) was estimated using the CR equation and placing ROIs over main lobe and side lobe regions respectively. In phantom experiments, ROIs were placed inside and outside anechoic cysts. For in vivo acquisitions, blood vessels were manually segmented using the Volume Segmenter tool in MATLAB.

Finally, the speckle signal-to-noise ratio (sSNR) was also estimated on the background of each image. The sSNR is given by \(sSNR=\left|\bar{A}\right|/\sqrt{var\left(A\right)}\), where A is the envelope level in some speckle region, the value \(\bar{A}\) is the mean and \(var(A)\) is the variance of the envelope in that region \cite{wagner_statistics_1983}. A higher sSNR indicates better speckle quality, with fully developed speckle approaching an sSNR value of 1.91 \cite{wagner_statistics_1983}.

\section{Results} \label{Results}

\subsection{Simulations} \label{section:simulations}
\begin{figure}[t]
    \centering
    \includegraphics[width=\linewidth]{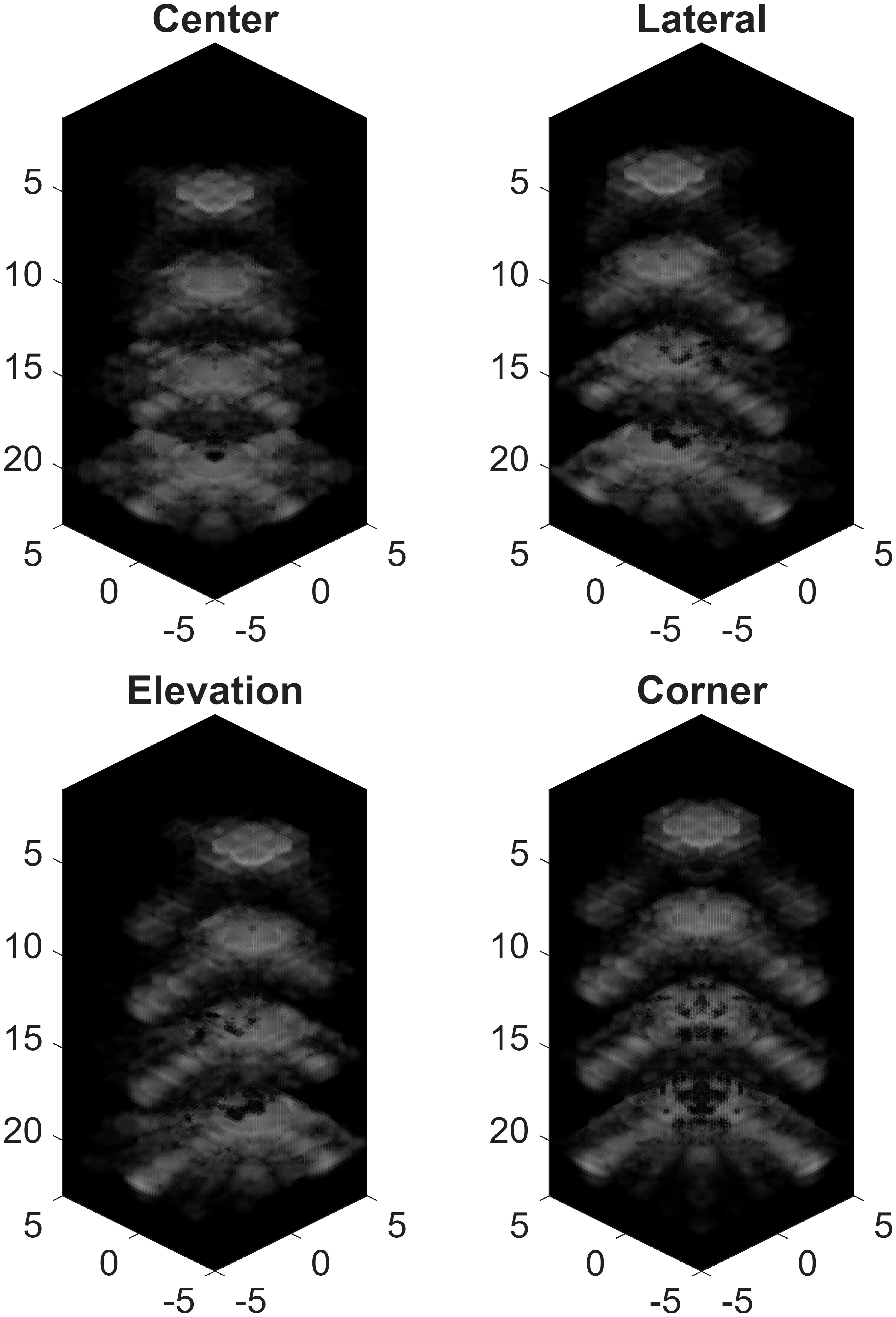}
    \caption{Simulated volumes of PSFs at different depths and lateral/elevational positions. All volumes in this Figure were beamformed with DAS and displayed with a dynamic range of 60 dB. Axis units are in mm.}
    \label{fig3:simulation_volumes}
\end{figure}

\begin{figure}[t]
    \centering
    \includegraphics[width=\linewidth]{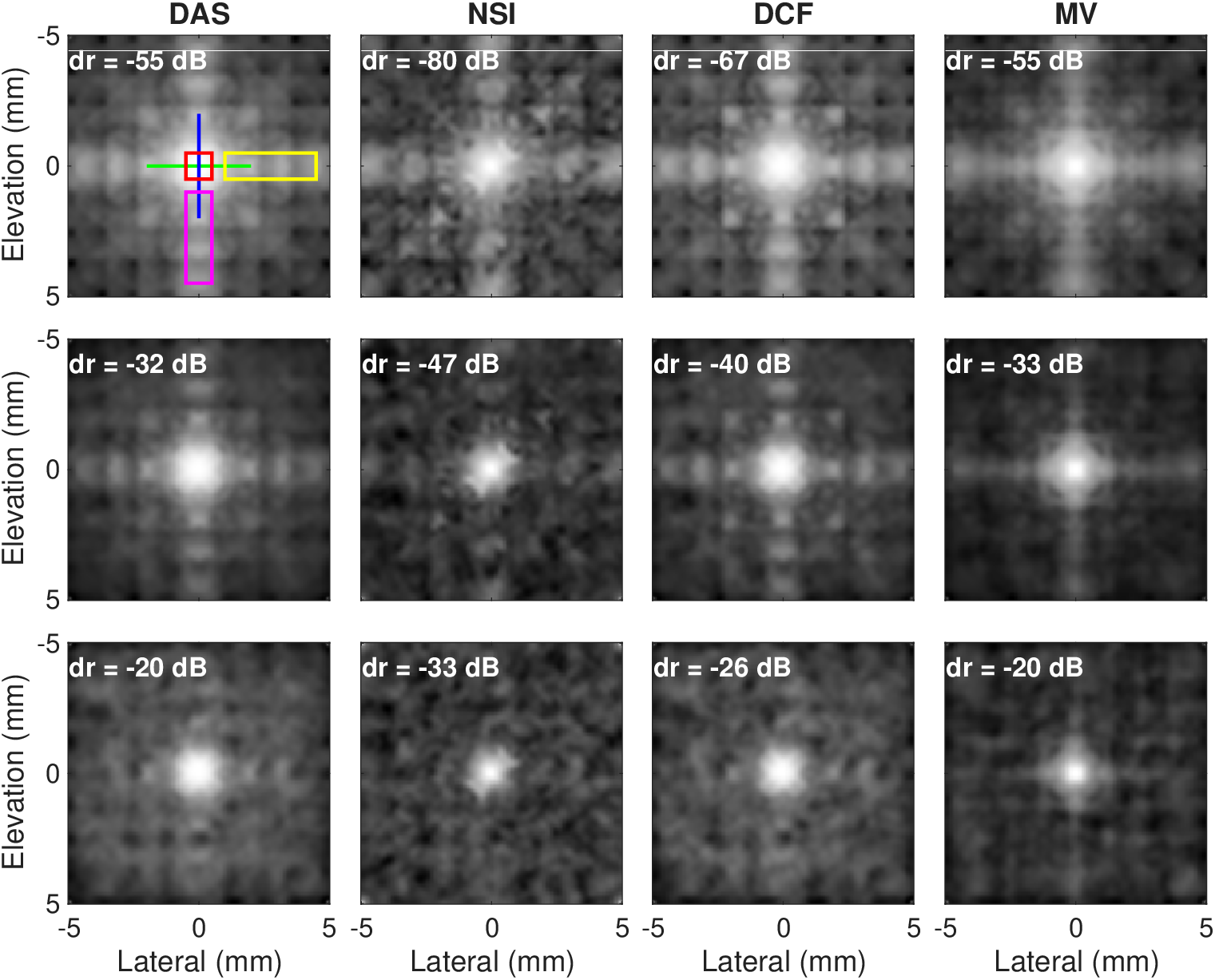}
    \caption{Depth slices at 15 mm for noiseless (top), noise amplitude 1e-3 (center), and noise amplitude 5e-3 (bottom) simulations. Slices were averaged axially with depth 1 mm for display. Illustrated ROIs were used for quality metric estimation. Note that while ROIs are displayed here in 2D for easier visualization, they also represent a depth of 1 mm in/out of the page. The dynamic range of each image was manually adjusted to maintain similar background brightness for all noise levels.}
    \label{fig4:simulation_depthSlices_noise}
\end{figure}

\begin{figure}[t]
    \centering
    \includegraphics[width=\linewidth]{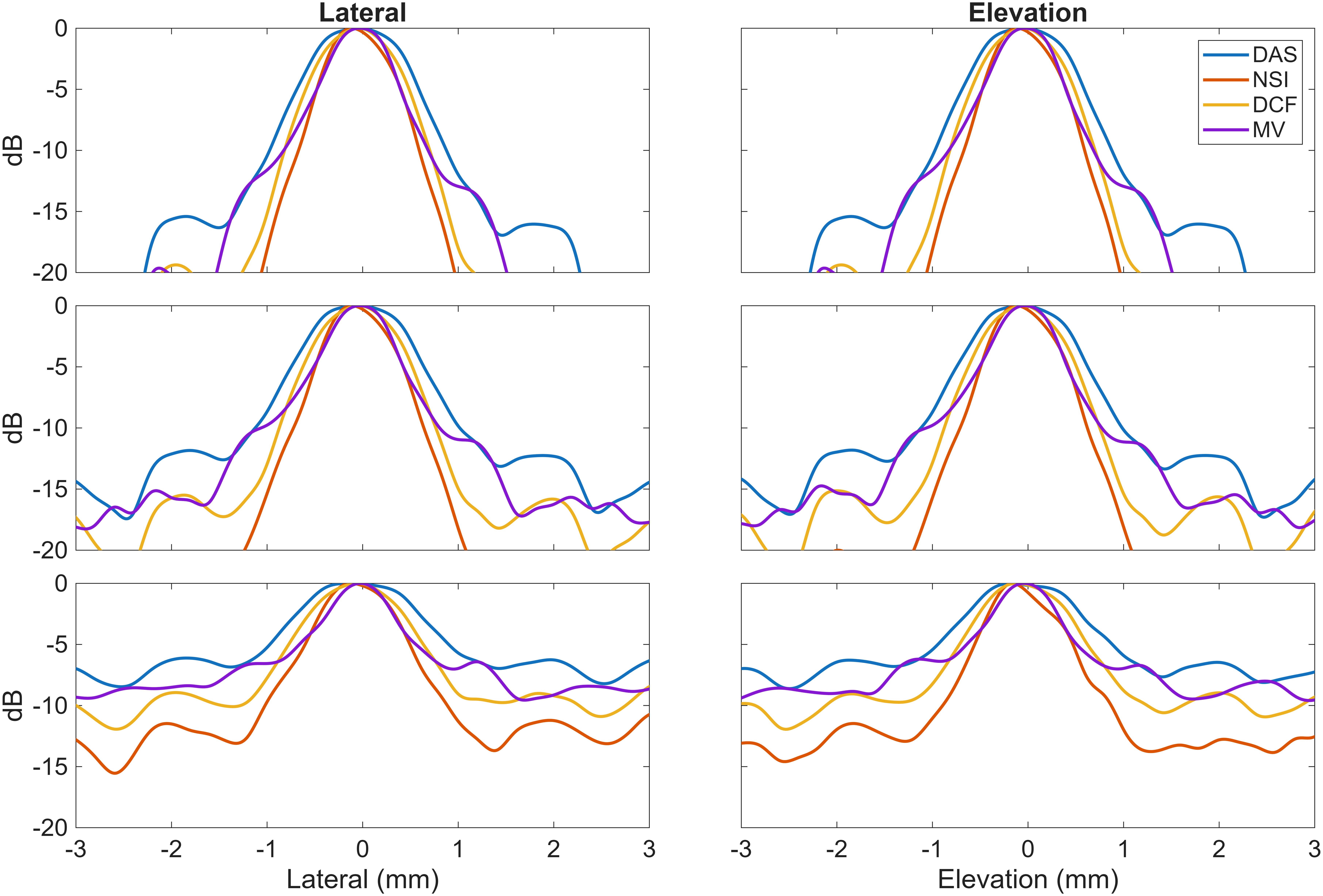}
    \caption{Simulated lateral and elevational profiles from the slices in Figure \ref{fig4:simulation_depthSlices_noise}.}
    \label{fig5:simulation_profiles}
\end{figure}

Volumetric B-mode images of points at different depths and lateral/elevational positions are displayed in Figure \ref{fig3:simulation_volumes} from the noiseless simulation. Depth slices at 15 mm of the simulated point-spread functions are displayed in Figure \ref{fig4:simulation_depthSlices_noise} for all three noise levels. Lateral and elevational profiles for each coupling number, beamformer, and noise level are illustrated in Figure \ref{fig5:simulation_profiles}. Quality metrics are given in Table \ref{tab1:kwave_quality_metrics} as mean \(\pm\) standard deviation over all point locations. 

We observed much greater variance between noise levels than between locations (see Table \ref{tab1:kwave_quality_metrics}). For the noiseless simulation, FWHM and MLSLR values are nearly identical for lateral and elevational estimates, reflecting the symmetry of the simulation. Increased noise levels resulted in both larger FWHM estimates and lower MLSLR values for every beamformer. 

For resolution, in the noiseless simulation, NSI had the lowest mean FWHM value of 0.85 mm in both directions. However, the noiseless FWHM estimate for DCF was within one standard deviation from NSI at 0.94 dB in both directions. For the medium noise case, the mean FWHM estimates of NSI and DCF are nearly identical in both directions, all between 0.95 and 0.97 mm. For the high noise case, DCF had the lowest mean FWHM estimates at 1.19 mm and 1.21 mm for lateral and elevational directions respectively. However, both of these values are within a standard deviation of NSI FWHM values, at 1.32 mm and 1.33 mm respectively. The mean FWHM estimates of MV were an improvement over DAS for all noise levels, but still slightly worse than both NSI and DCF. In resolution, the DCF beamformer appeared to be the most resilient against noise, having similar FWHM estimates between the noiseless and medium noise case, and the lowest FWHM estimates at the highest noise level. 

For side lobe levels, NSI had consistently the highest mean MLSLR estimates across all noise levels. Lower side lobes with NSI can also be observed in lateral and elevational profiles (Figure \ref{fig5:simulation_profiles}). The next highest MLSLR estimates came from DCF, but were more than a standard deviation lower than NSI for all noise levels. The MLSLR estimates for DAS and MV were similar to each other for both directions and all noise levels. 

\begin{table*}[ht]
	\centering
	\begin{tabular}{|c|c|c|c|c|}
		\hline
		Beamformer & FWHM lat (mm) & FWHM ele (mm) & MLSLR lat (dB) & MLSLR ele (dB) \\
		\hline
		DAS Noiseless & 1.36 \(\pm\) 0.08 & 1.36 \(\pm\) 0.08 & 9.65 \(\pm\) 0.76 & 9.65 \(\pm\) 0.76\\
		NSI Noiseless & 0.85 \(\pm\) 0.14 & 0.85 \(\pm\) 0.14 & 12.89 \(\pm\) 1.21 & 12.89 \(\pm\) 1.21\\
		DCF Noiseless & 0.94 \(\pm\) 0.27 & 0.94 \(\pm\) 0.27 & 10.93 \(\pm\) 0.59 & 10.93 \(\pm\) 0.60\\
		MV Noiseless & 1.04 \(\pm\) 0.27 & 1.04 \(\pm\) 0.27 & 9.76 \(\pm\) 0.82 & 9.76 \(\pm\) 0.82\\
		\hline
		DAS Noise 1e-3 & 1.53 \(\pm\) 0.14 & 1.52 \(\pm\) 0.14 & 7.45 \(\pm\) 0.75 & 7.49 \(\pm\) 0.73\\
		NSI Noise 1e-3 & 0.96 \(\pm\) 0.10 & 0.95 \(\pm\) 0.08 & 9.83 \(\pm\) 0.85 & 9.92 \(\pm\) 0.99\\
		DCF Noise 1e-3 & 0.97 \(\pm\) 0.29 & 0.97 \(\pm\) 0.29 & 8.64 \(\pm\) 0.64 & 8.69 \(\pm\) 0.64\\
		MV Noise 1e-3 & 1.19 \(\pm\) 0.37 & 1.17 \(\pm\) 0.36 & 7.37 \(\pm\) 0.91 & 7.42 \(\pm\) 0.90\\
		\hline
		DAS Noise 5e-3 & 2.43 \(\pm\) 0.51 & 2.31 \(\pm\) 0.30 & 3.84 \(\pm\) 0.44 & 3.83 \(\pm\) 0.42\\
		NSI Noise 5e-3 & 1.32 \(\pm\) 0.29 & 1.33 \(\pm\) 0.29 & 5.30 \(\pm\) 0.80 & 5.33 \(\pm\) 0.75\\
		DCF Noise 5e-3 & 1.19 \(\pm\) 0.37 & 1.21 \(\pm\) 0.36 & 4.74 \(\pm\) 0.58 & 4.74 \(\pm\) 0.57\\
		MV Noise 5e-3 & 1.95 \(\pm\) 0.53 & 1.94 \(\pm\) 0.54 & 3.70 \(\pm\) 0.64 & 3.69 \(\pm\) 0.63\\
		\hline
	\end{tabular}
	\caption{Quality metrics for the simulated aperture, averaged over all 12 point locations displayed in Figure \ref{fig3:simulation_volumes}. Metrics are given as mean \(\pm\) standard deviation.}
	\label{tab1:kwave_quality_metrics}
\end{table*}

\subsection{Virtual Large Aperture}
\begin{figure}
    \centering
    \includegraphics[width=\linewidth]{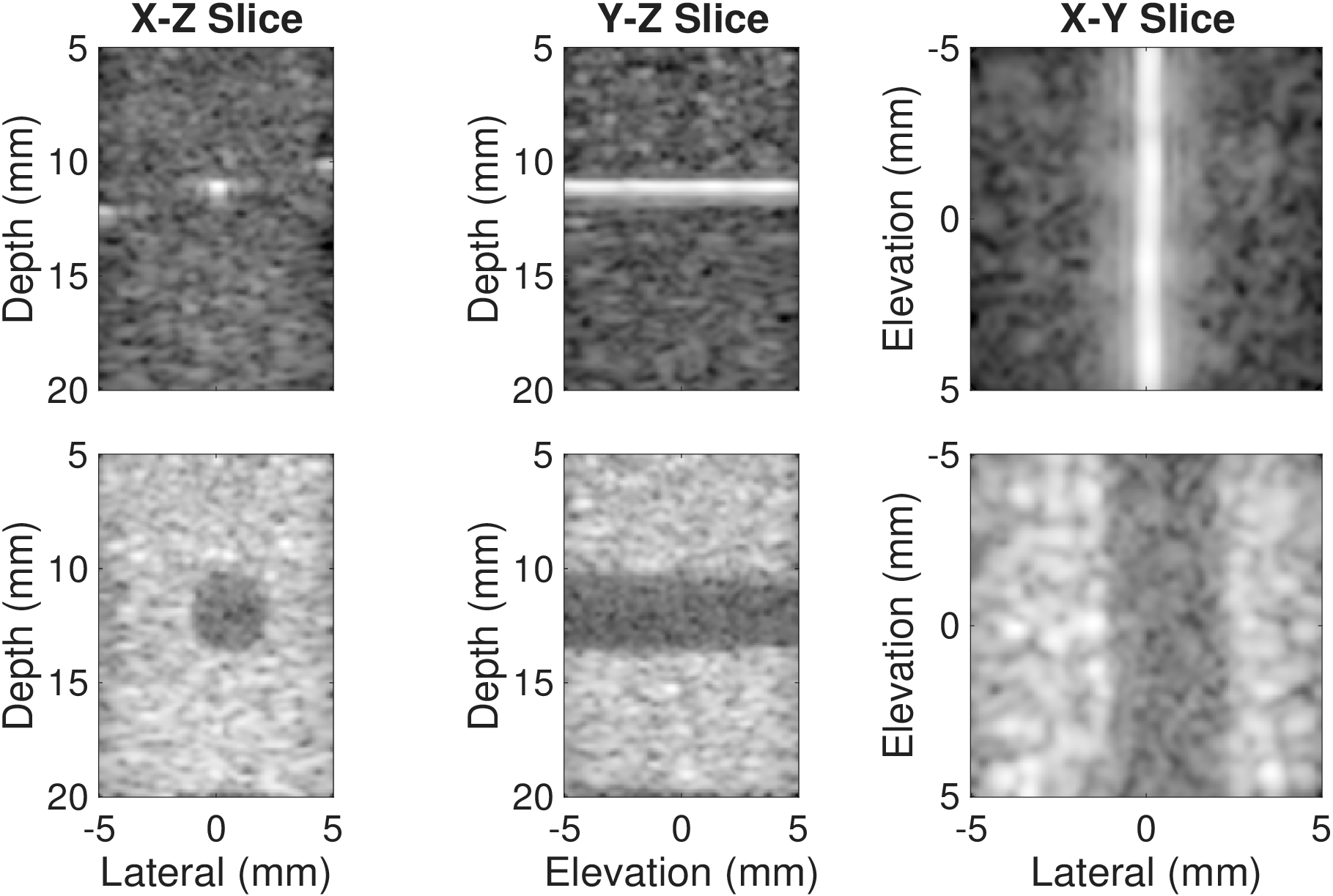}
    \caption{B-mode images of the wire and cyst in the ATS phantom using only the Vermon array with no element coupling.}
    \label{fig6:bmode_vermon_only}
\end{figure}

\begin{figure}
    \centering
    \includegraphics[width=\linewidth]{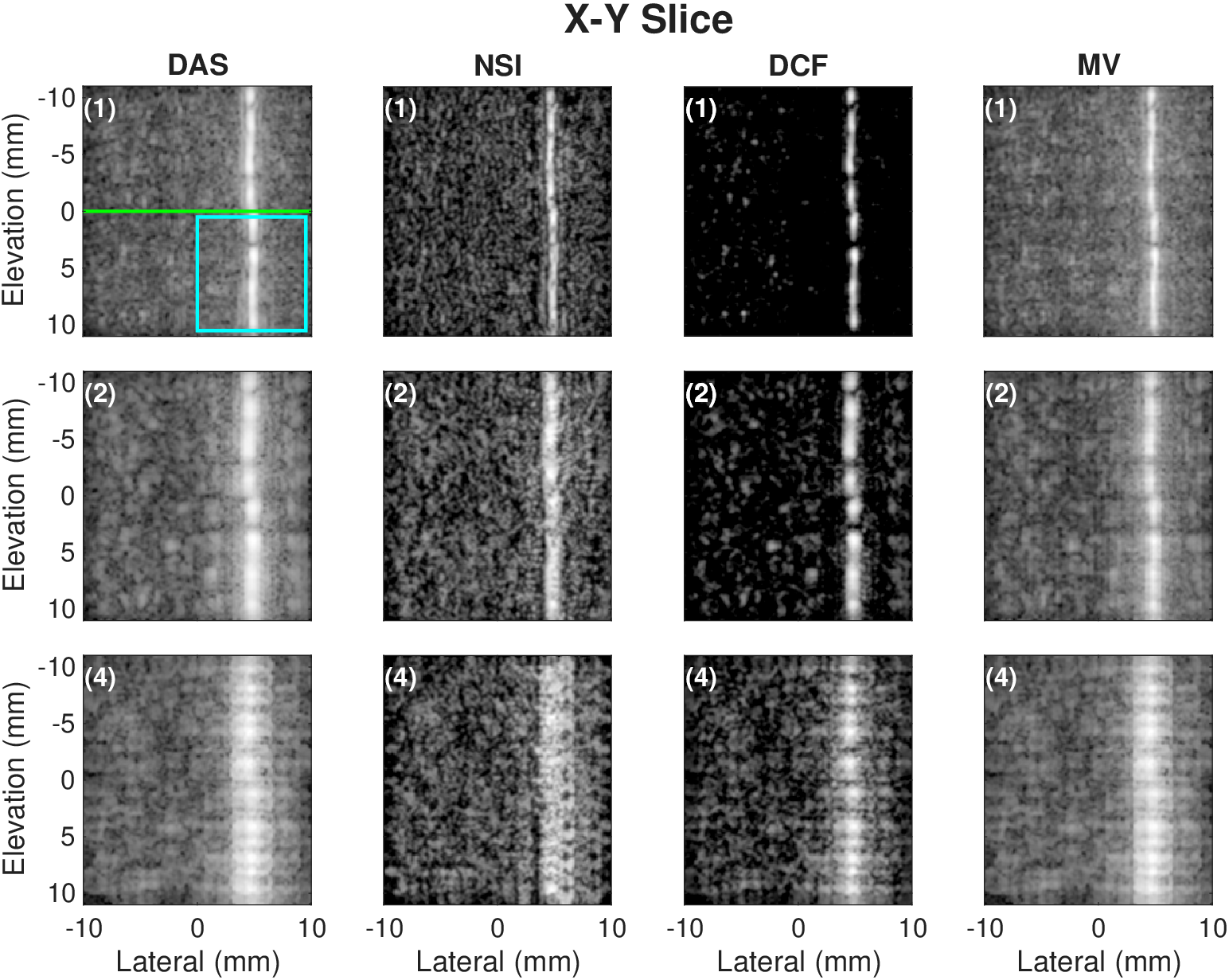}
    \caption{Lateral/elevational slices of the wire phantom from the virtual large aperture at a depth of 9.6 mm. All images are displayed with a dynamic range of 60 dB. The coupling number is indicated in the top left corner of each panel. Top to bottom goes coupling numbers 1, 2, and 4. The green line illustrates the region for the lateral profiles and FWHM estimates. The cyan box indicates the equivalent FOV from the Vermon array by itself.}
    \label{fig7:wire_xyslices}
\end{figure}

\begin{figure}
    \centering
    \includegraphics[width=\linewidth]{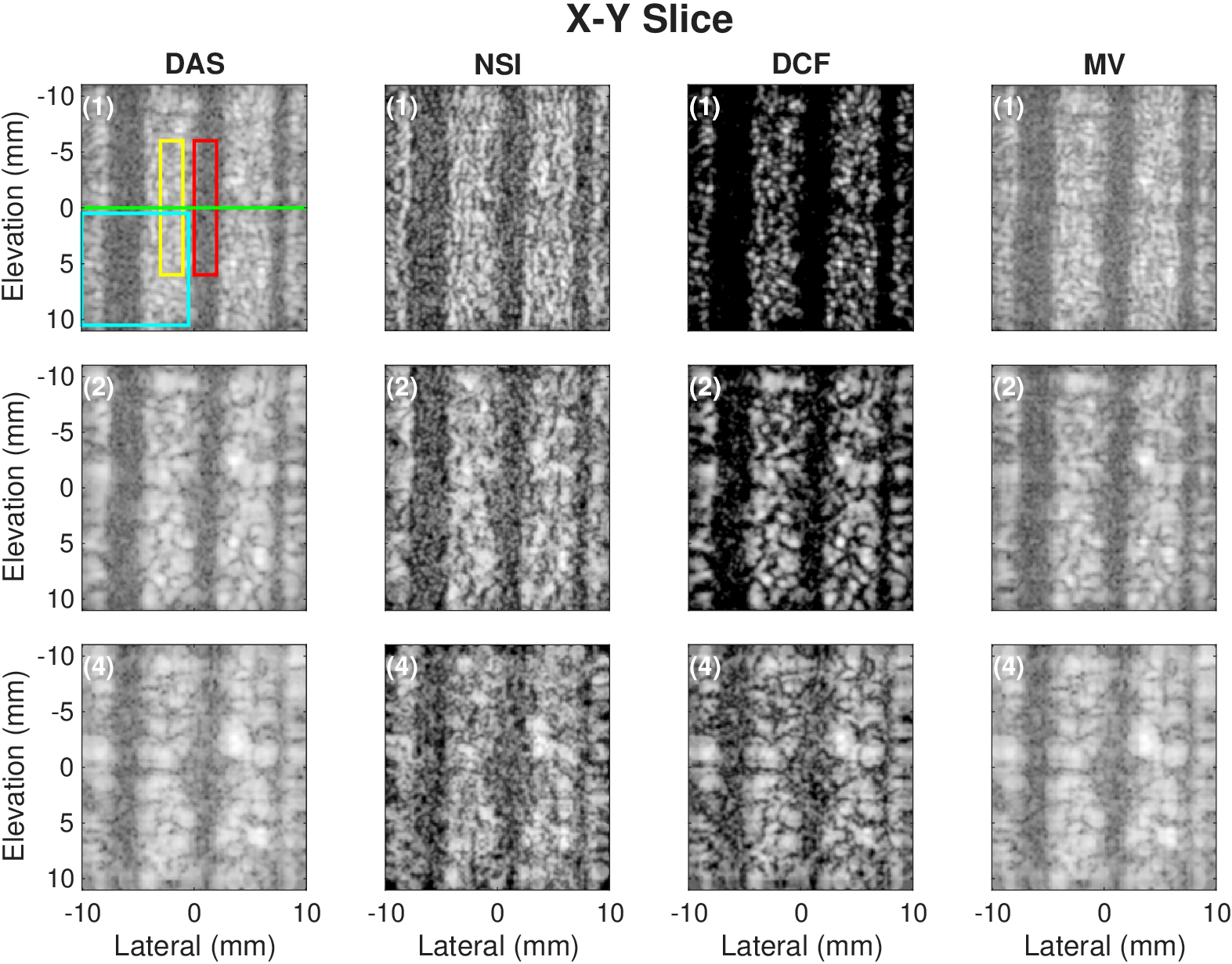}
    \caption{Lateral/elevational slices of the cyst phantom from the virtual large aperture at a depth of 10.4 mm. All images are displayed with a dynamic range of 60 dB. The coupling number is indicated in the top left corner of each panel. Top to bottom goes coupling numbers 1, 2, and 4. The red and yellow boxes illustrate the ROIs for contrast metrics. The green line indicates the lateral profile for Figure \ref{fig11:lateral_profiles}. The cyan box indicates the equivalent FOV from the Vermon array by itself.}
    \label{fig8:cyst_xyslices}
\end{figure}

\begin{figure*}
    \centering
    \includegraphics[width=\linewidth]{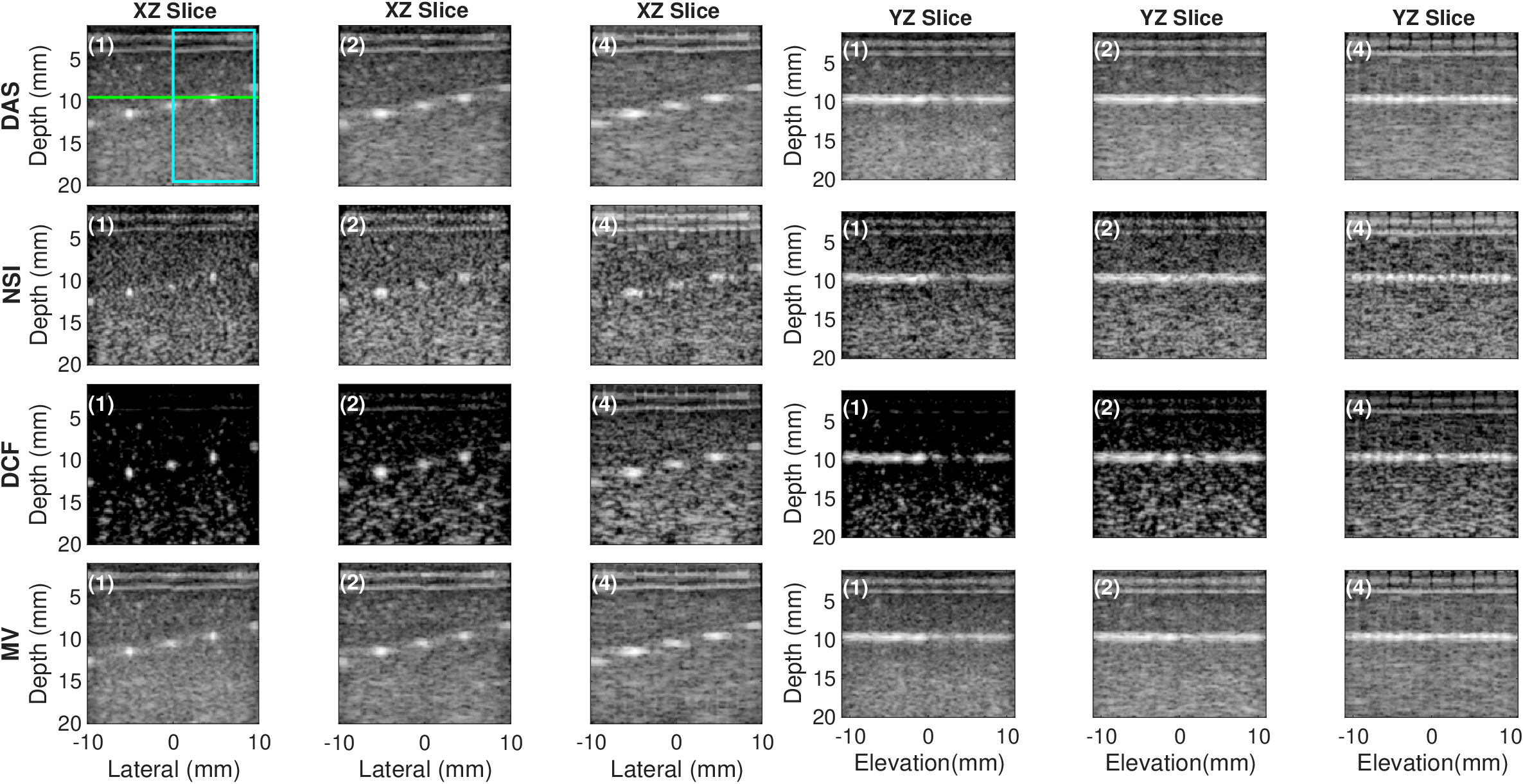}
    \caption{Wire phantom image slices from the virtual large aperture. Top to bottom goes DAS, NSI, DCF, and MV. The coupling numbers are indicated in the top left corner of each panel. The first three columns represent coupling numbers 1, 2, and 4 of the lateral slice. The second three columns represent coupling numbers 1, 2, and 4 of the elevational slice. All images are displayed with a 60 dB dynamic range. The green line represents the lateral profile for Figure \ref{fig11:lateral_profiles}. The cyan box indicates the equivalent FOV from the Vermon array by itself.}
    \label{fig9:wire_xz_yz_slices}
\end{figure*}

\begin{figure*}
    \centering
    \includegraphics[width=\linewidth]{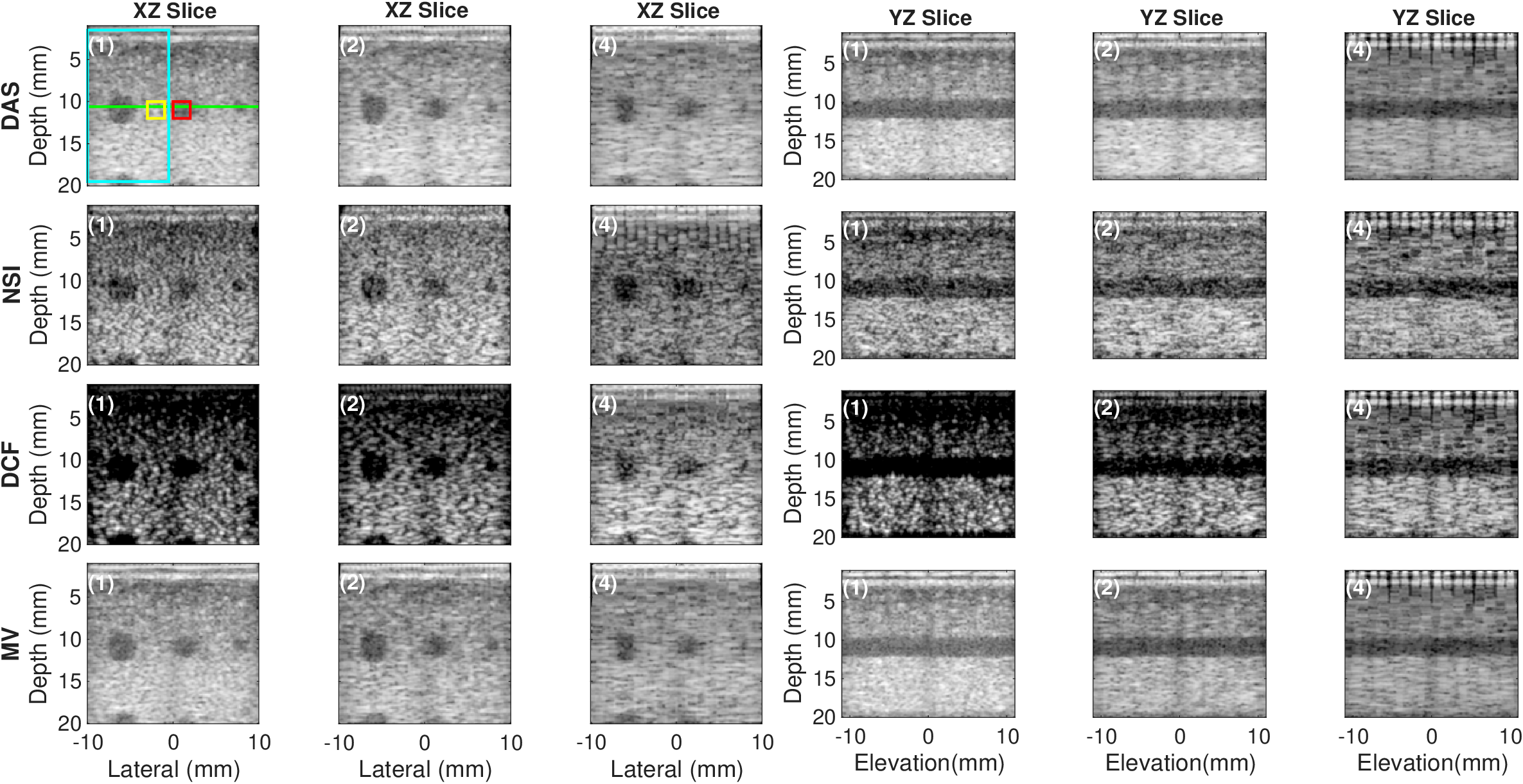}
    \caption{Cyst phantom image slices from the virtual large aperture. Top to bottom goes DAS, NSI, DCF, and MV. The coupling numbers are indicated in the top left corner of each panel. The first three columns represent coupling numbers 1, 2, and 4 of the lateral slice. The second three columns represent coupling numbers 1, 2, and 4 of the elevational slice. All images are displayed with a 60 dB dynamic range. The green line represents the lateral profile for Figure \ref{fig11:lateral_profiles}. The red and yellow boxes indicate the ROIs for contrast metrics. The cyan box indicates the equivalent FOV from the Vermon array by itself.}
    \label{fig10:cyst_xz_yz_slices}
\end{figure*}

\begin{figure}
    \centering
    \includegraphics[width=\linewidth]{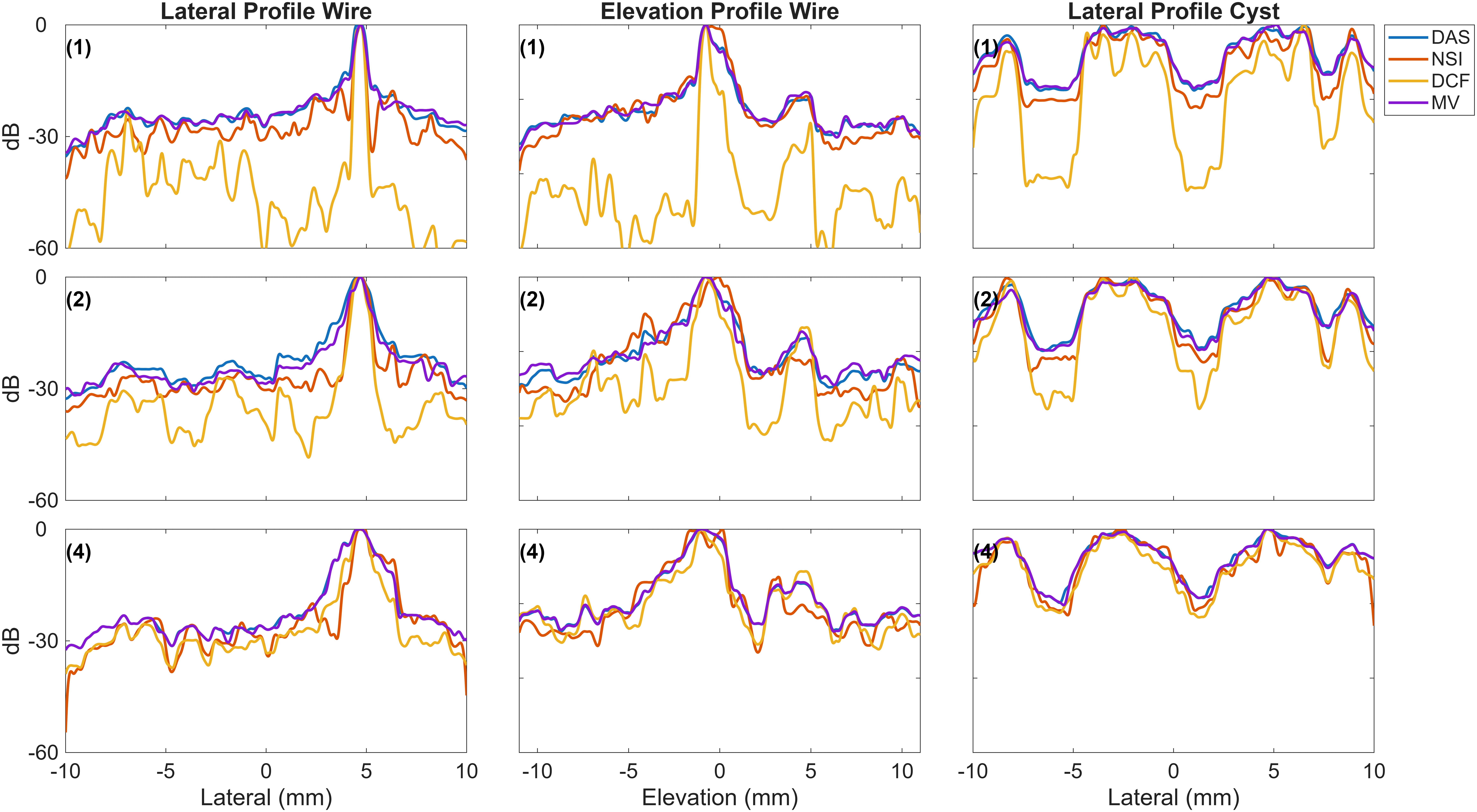}
    \caption{Lateral and elevation profiles from the virtual large aperture, along the green lines in Figures \ref{fig7:wire_xyslices}-\ref{fig10:cyst_xz_yz_slices}. The first column is lateral profiles of a wire target denoted by the green line in Figure \ref{fig9:wire_xz_yz_slices}, top left. The second column is elevational profiles of the same wire target after the virtual aperture was rotated. An extra side lobe is visible in the elevation profile because of the panel separation. The final column is lateral profiles of anechoic regions denoted by the green line in Figure \ref{fig10:cyst_xz_yz_slices}, top left. Top to bottom goes coupling numbers of 1, 2, and 4.}
    \label{fig11:lateral_profiles}
\end{figure}

Figure \ref{fig6:bmode_vermon_only} displays B-mode images using the Vermon array by itself for FOV comparison. Figs. \ref{fig7:wire_xyslices}-\ref{fig10:cyst_xz_yz_slices} display B-mode images acquired with the virtual large aperture. First, Figs. \ref{fig7:wire_xyslices}-\ref{fig8:cyst_xyslices} display the constant depth slices of wire and cyst phantoms respectively, with each coupling number and beamformer. The exact depths were 9.6 mm for the wires, to align with an off-center wire, and 10.4 mm for the cysts, to align with the center of all three cysts. We note that while we only display slices, quality metrics were evaluated on full volume data. Illustrated ROIs are cross sections of the true volumetric ROIs. Then, Figs. \ref{fig9:wire_xz_yz_slices}-\ref{fig10:cyst_xz_yz_slices} display the lateral and elevational slices of the wire and cyst phantoms, again with each coupling number. Fig. \ref{fig11:lateral_profiles} displays lateral profiles for the wire targets and cyst regions, denoted by the green line in Figs. \ref{fig9:wire_xz_yz_slices}-\ref{fig10:cyst_xz_yz_slices}. Finally, Table \ref{tab3:vla_quality_metrics} gives the quality metrics for each beamformer and coupling number. 

\begin{table}[ht]
    \begin{adjustbox}{width=\linewidth,center}
    \begin{tabular}{|c|c|c|c|}
        \hline
        Coupling Number & Element Count & Width (\(\lambda\)) & F\# \\
        \hline
        1 & 4096 & 1.48 & 1.70\\
        2 & 1024 & 3.10 & 3.51\\
        4 & 256 & 6.32 & 7.16\\
        \hline
    \end{tabular}
    \end{adjustbox}
    \caption{Element counts, widths, and minimum F-numbers (Eq. \ref{eq:minimum_fnumber}) for each coupling number on the virtual large aperture.}
    \label{tab2:vla_counts_widths}
\end{table}

With a coupling number of two, the total element count of the virtual aperture was the same as the Vermon matrix array, meaning the aperture size and FOV doubled in both dimensions with no increase to the element count. We observed a degradation to resolution with increasing coupling numbers, with mean lateral FWHM estimates at a depth of 9.6 mm going from 0.78 mm with no coupling to 1.98 mm with coupling by four, both with DAS. The mean elevational FWHM estimates also increased from 1.16 mm without coupling to 2.23 mm with coupling by 4 using DAS. The elevational FWHM estimates are consistently worse than corresponding lateral FWHM estimates for each beamformer. This is especially apparent with NSI, comparing its 0.46 mm FWHM laterally to an estimated FWHM of 1.84 mm in elevation without coupling. With coupling by two, lateral FWHM estimates improved from 1.34 mm with DAS to 0.78 mm with DCF beamforming, while elevational FWHM improved from 1.63 mm with DAS to 1.18 mm with DCF. Even with coupling by 4, DCF maintained similar FWHM estimates at 0.94 mm laterally and 1.13 mm in elevation. 

As for contrast metrics, both the contrast ratio and the gCNR decreased with higher coupling numbers. The most extreme case for contrast ratio came from the DCF beamformer which went from -30.6 dB without coupling to -10.0 dB with coupling by 4. With gCNR, the DAS beamformer went from 0.70 without coupling to 0.63 coupling by 2 and 0.56 coupling by 4. The CNR stayed fairly consistent for each beamformer across coupling numbers. The mean sSNR values increased for each beamformer with higher coupling numbers. The sSNR increase is likely due to increased size of the resolution cell: more scatterers are included in a wider resolution cell leading to more fully developed speckle. 

The DCF beamformer had the highest contrast ratio for coupling by 1 and 2, while DAS, MV, and NSI were all relatively similar being within a standard deviation of each other for coupling by 1 and 2. Each beamformer produced similar contrast metrics for coupling by 4. The DAS and MV beamformers had the highest CNR, gCNR, and sSNR across all coupling numbers. With CNR, NSI was higher than DCF without coupling. With coupling by 2 and 4, NSI and DCF had CNR values within one standard deviation of each other. With gCNR, NSI and DCF had very similar values for each coupling number. DCF had the lowest sSNR values for coupling by 1 and 2, while NSI and DCF were similarly low for coupling by 4. 

As a final observation, there are ``dead zone" artifacts visible in the elevational slices of coupling by 4 (Figs. \ref{fig9:wire_xz_yz_slices}-\ref{fig10:cyst_xz_yz_slices}, final column). Several pixels near the transducer face are not visible to any elements due to narrow directivity and panel separation on the Vermon array in the elevational direction. Therefore, these artifacts are more prevalent in elevation than laterally.

\begin{table*}[ht]
	\centering
	\begin{tabular}{|c|c|c|c|c|c|c|}
		\hline
		Beamformer & FWHM lat (mm) & FWHM ele (mm) & Contrast (dB) & CNR & gCNR & sSNR \\
		\hline
		DAS C1 & 0.78 \(\pm\) 0.09 & 1.16 \(\pm\) 0.21 & -12.0 \(\pm\) 3.2 & -1.08 \(\pm\) 0.19 & 0.70 \(\pm\) 0.12 & 1.52 \(\pm\) 0.19\\
		NSI C1 & 0.46 \(\pm\) 0.02 & 1.84 \(\pm\) 0.53 & -12.9 \(\pm\) 4.1 & -0.61 \(\pm\) 0.13 & 0.39 \(\pm\) 0.11 & 0.84 \(\pm\) 0.08\\
		DCF C1 & 0.49 \(\pm\) 0.02 & 0.45 \(\pm\) 0.08 & -30.6 \(\pm\) 6.7 & -0.34 \(\pm\) 0.09 & 0.41 \(\pm\) 0.12 & 0.35 \(\pm\) 0.09\\
		MV C1 & 0.77 \(\pm\) 0.19 & 1.20 \(\pm\) 0.39 & -11.3 \(\pm\) 3.1 & -1.00 \(\pm\) 0.17 & 0.66 \(\pm\) 0.13 & 1.45 \(\pm\) 0.18\\
		\hline
		DAS C2 & 1.34 \(\pm\) 0.06 & 1.63 \(\pm\) 0.45 & -9.8 \(\pm\) 4.7 & -1.07 \(\pm\) 0.30 & 0.63 \(\pm\) 0.19 & 1.75 \(\pm\) 0.01\\
		NSI C2 & 0.90 \(\pm\) 0.09 & 1.60 \(\pm\) 0.29 & -11.5 \(\pm\) 5.9 & -0.61 \(\pm\) 0.19 & 0.38 \(\pm\) 0.17 & 0.93 \(\pm\) 0.02\\
		DCF C2 & 0.78 \(\pm\) 0.10 & 1.18 \(\pm\) 0.34 & -17.3 \(\pm\) 8.5 & -0.50 \(\pm\) 0.16 & 0.38 \(\pm\) 0.20 & 0.60 \(\pm\) 0.10\\
		MV C2 & 1.03 \(\pm\) 0.11 & 1.26 \(\pm\) 0.40 & -10.1 \(\pm\) 4.6 & -1.03 \(\pm\) 0.26 & 0.62 \(\pm\) 0.18 & 1.64 \(\pm\) 0.04\\
		\hline
		DAS C4 & 1.98 \(\pm\) 0.25 & 2.23 \(\pm\) 0.19 & -8.1 \(\pm\) 3.4 & -0.99 \(\pm\) 0.30 & 0.56 \(\pm\) 0.16 & 1.86 \(\pm\) 0.03\\
		NSI C4 & 1.18 \(\pm\) 0.32 & 1.80 \(\pm\) 0.79 & -9.0 \(\pm\) 4.8 & -0.53 \(\pm\) 0.21 & 0.33 \(\pm\) 0.14 & 0.95 \(\pm\) 0.02\\
		DCF C4 & 0.94 \(\pm\) 0.04 & 1.13 \(\pm\) 0.12 & -10.0 \(\pm\) 4.3 & -0.58 \(\pm\) 0.19 & 0.35 \(\pm\) 0.14 & 0.93 \(\pm\) 0.06\\
		MV C4 & 1.95 \(\pm\) 0.27 & 2.23 \(\pm\) 0.18 & -8.3 \(\pm\) 3.4 & -1.00 \(\pm\) 0.30 & 0.58 \(\pm\) 0.15 & 1.83 \(\pm\) 0.01\\
		\hline
	\end{tabular}
	\caption{Quality metrics for the virtual aperture, estimated using the ROIs illustrated in Figures \ref{fig7:wire_xyslices}-\ref{fig10:cyst_xz_yz_slices}. ``C1", ``C2", and ``C4" denote the coupling number. Metrics are given as mean \(\pm\) standard deviation over three repeated measurements.}
	\label{tab3:vla_quality_metrics}
\end{table*}

\subsection{In vivo}
\begin{figure}
    \centering
    \includegraphics[width=\linewidth]{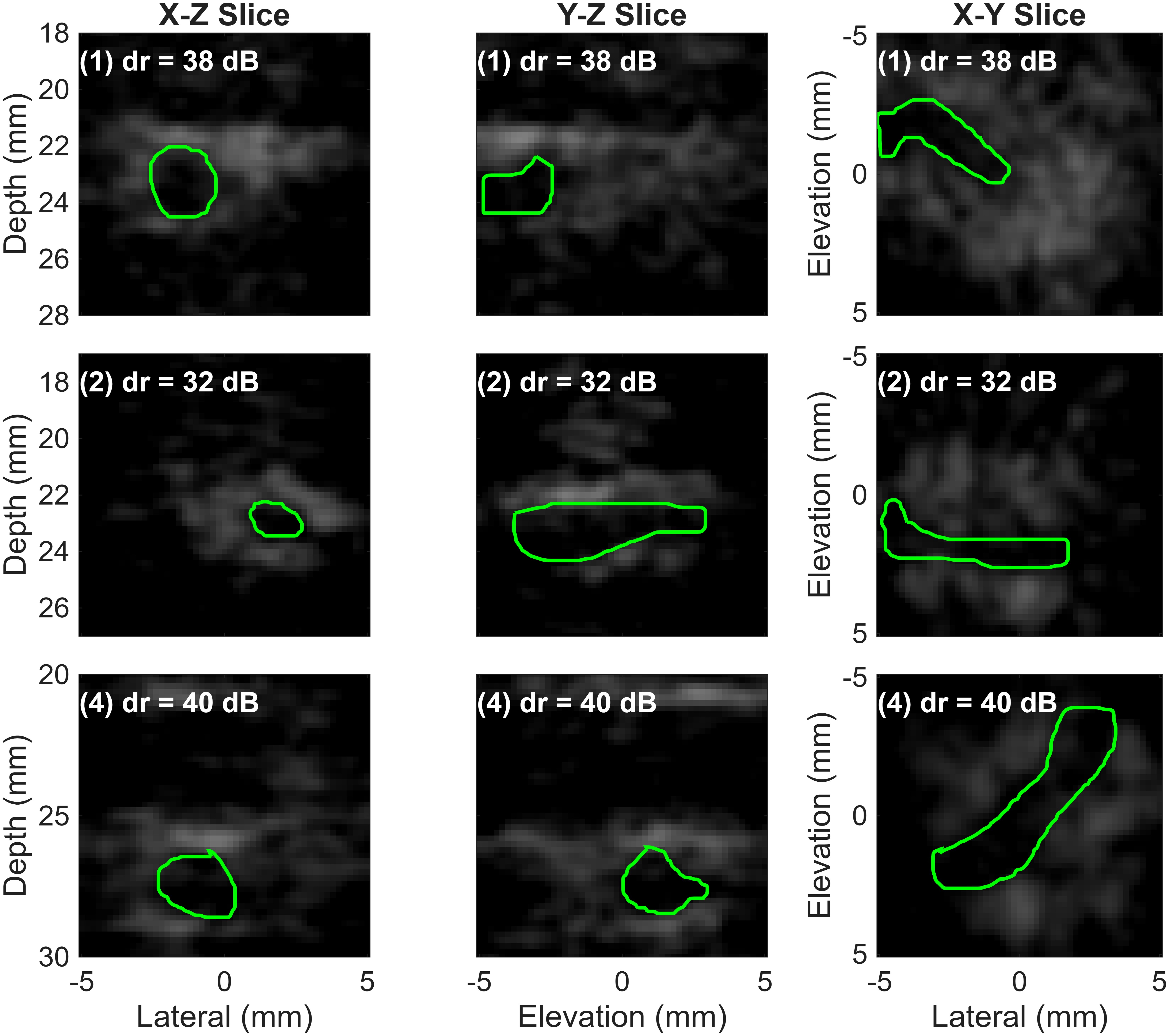}
    \caption{B-mode slices of blood vessels in a rabbit liver, beamformed using DAS. Top to bottom goes coupling numbers of 1, 2, and 4, and they are also indicated in the top left corner of each panel. The dynamic range was manually adjusted on each image to highlight the vessel boundaries. The manually drawn segments used for contrast metrics are also displayed as green boundaries.}
    \label{fig12:bmode_invivo_coupling_slices}
\end{figure}

\begin{figure}
    \centering
    \includegraphics[width=\linewidth]{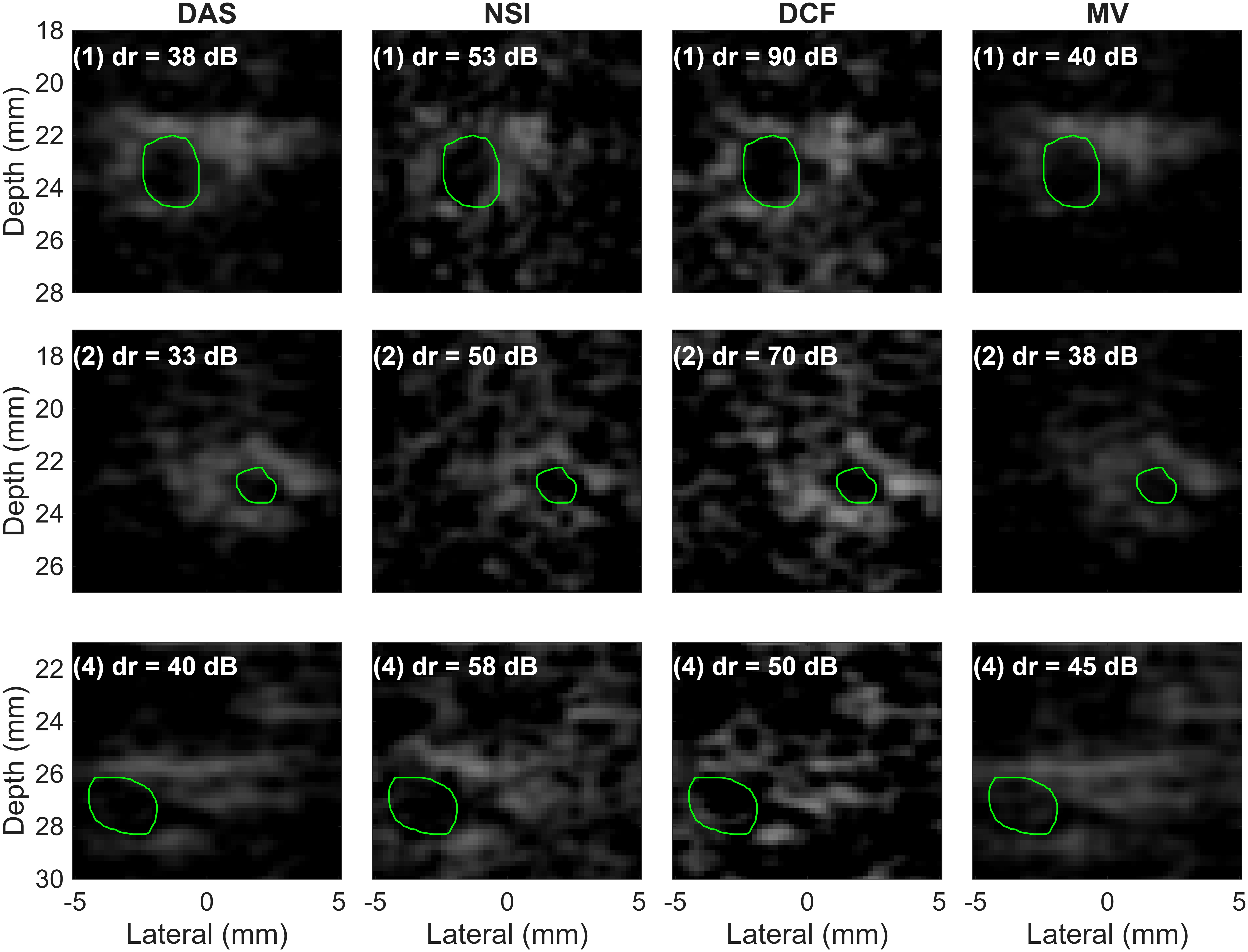}
    \caption{B-mode lateral slices of blood vessels in a rabbit liver, using all four beamformers. Left to right goes DAS, NSI, DCF, MV. Top to bottom goes coupling numbers of 1, 2, and 4, and they are also indicated in the top left corner of each panel. The dynamic range was adjusted on each image to highlight the vessel boundaries. The manually drawn segments used for contrast metrics are also displayed as green boundaries.}
    \label{fig13:bmode_invivo_coupling_beamforming}
\end{figure}

Figure \ref{fig12:bmode_invivo_coupling_slices} displays example lateral, elevational, and depth slices of blood vessels in the rabbit liver, all processed with DAS beamforming. Figure \ref{fig13:bmode_invivo_coupling_beamforming} displays only lateral slices for each coupling number and beamformer. Table \ref{tab4:invivo_quality_metrics} lists quality metrics averaged over nine repeated measurements for each beamformer and coupling number. 

The lateral autocorrelation FWHM estimates are consistently higher than corresponding elevational FWHM estimates for each beamformer. Much like the phantom experiment with the virtual aperture, increased coupling resulted in increased FWHM estimates in both directions for all beamformers. The DAS beamformer went from mean lateral and elevation FWHM values of 1.81 mm and 1.48 mm without coupling to 2.03 mm and 1.94 mm coupling by 2, and 2.59 mm and 2.24 mm coupling by 4. NSI and DCF beamformers have the lowest FWHM values across all coupling numbers. DCF only seems to exceed the resolution performance of NSI in elevational FWHM without coupling. The MV and DAS beamformers have similar FWHM estimates to each other, and higher FWHM estimates than NSI and DCF, in both directions. 

Like the phantom experiments with the virtual aperture, DCF had the lowest contrast ratio estimates across all coupling numbers. However, the difference between DCF and other beamformers in this metric is only larger than a standard deviation for the no coupling case. The CNR metrics were all very low and highly variable for each beamformer and coupling number. DAS and MV appear to have higher gCNR than NSI or DCF, although all differences in the means are smaller than the standard deviations of DAS and MV. DCF experiences increased sSNR with higher coupling numbers. The other beamformers stay fairly consistent (within a standard deviation) across coupling numbers. Like the phantom experiment with the virtual aperture, DAS and MV have the highest sSNR metrics for each coupling number. Then, NSI has the next highest sSNR, and DCF has the lowest sSNR metrics. 

\begin{table*}[ht]
	\centering
	\begin{tabular}{|c|c|c|c|c|c|c|}
		\hline
		Beamformer & FWHM lat (mm) & FWHM ele (mm) & Contrast (dB) & CNR & gCNR & sSNR \\
		\hline
		DAS C1 & 1.81 \(\pm\) 0.56 & 1.48 \(\pm\) 0.30 & -3.33 \(\pm\) 2.22 & -0.35 \(\pm\) 0.21 & 0.24 \(\pm\) 0.13 & 1.36 \(\pm\) 0.22\\
		NSI C1 & 0.43 \(\pm\) 0.05 & 0.56 \(\pm\) 0.10 & -3.09 \(\pm\) 2.41 & -0.19 \(\pm\) 0.15 & 0.15 \(\pm\) 0.07 & 0.77 \(\pm\) 0.09\\
		DCF C1 & 0.44 \(\pm\) 0.03 & 0.40 \(\pm\) 0.05 & -14.72 \(\pm\) 6.98 & -0.22 \(\pm\) 0.06 & 0.15 \(\pm\) 0.06 & 0.29 \(\pm\) 0.06\\
		MV C1 & 1.86 \(\pm\) 0.80 & 1.53 \(\pm\) 0.41 & -2.30 \(\pm\) 2.13 & -0.26 \(\pm\) 0.23 & 0.22 \(\pm\) 0.10 & 1.44 \(\pm\) 0.18\\
		\hline
		DAS C2 & 2.03 \(\pm\) 0.30 & 1.94 \(\pm\) 0.23 & -1.21 \(\pm\) 3.50 & -0.17 \(\pm\) 0.34 & 0.22 \(\pm\) 0.11 & 1.57 \(\pm\) 0.26\\
		NSI C2 & 0.53 \(\pm\) 0.19 & 0.55 \(\pm\) 0.18 & -1.12 \(\pm\) 4.23 & -0.07 \(\pm\) 0.25 & 0.16 \(\pm\) 0.07 & 0.83 \(\pm\) 0.09\\
		DCF C2 & 0.53 \(\pm\) 0.09 & 0.52 \(\pm\) 0.09 & -4.24 \(\pm\) 10.45 & -0.14 \(\pm\) 0.18 & 0.16 \(\pm\) 0.06 & 0.43 \(\pm\) 0.07\\
		MV C2 & 1.82 \(\pm\) 0.30 & 1.68 \(\pm\) 0.13 & -1.13 \(\pm\) 3.01 & -0.14 \(\pm\) 0.32 & 0.20 \(\pm\) 0.09 & 1.50 \(\pm\) 0.25\\
		\hline
		DAS C4 & 2.59 \(\pm\) 0.19 & 2.24 \(\pm\) 0.25 & -1.64 \(\pm\) 3.09 & -0.20 \(\pm\) 0.35 & 0.22 \(\pm\) 0.08 & 1.43 \(\pm\) 0.20\\
		NSI C4 & 0.74 \(\pm\) 0.11 & 0.70 \(\pm\) 0.11 & -1.76 \(\pm\) 2.93 & -0.11 \(\pm\) 0.18 & 0.13 \(\pm\) 0.05 & 0.80 \(\pm\) 0.11\\
		DCF C4 & 0.63 \(\pm\) 0.18 & 0.61 \(\pm\) 0.15 & -3.75 \(\pm\) 4.91 & -0.17 \(\pm\) 0.20 & 0.15 \(\pm\) 0.08 & 0.62 \(\pm\) 0.06\\
		MV C4 & 2.55 \(\pm\) 0.17 & 2.05 \(\pm\) 0.20 & -1.31 \(\pm\) 2.66 & -0.16 \(\pm\) 0.30 & 0.19 \(\pm\) 0.09 & 1.40 \(\pm\) 0.20\\
		\hline
	\end{tabular}
	\caption{Quality metrics for the in vivo study. ``C1," ``C2," and ``C4" indicate the coupling number. Metrics are given as mean \(\pm\) standard deviation over a total of nine repeated estimates.}
	\label{tab4:invivo_quality_metrics}
\end{table*}

\section{Discussion}
Our results demonstrate how larger elements can lead to larger apertures, as well as highlight trade-offs between different beamformers that could be used to maintain resolution. We observed that even with a coupling number of 4, NSI and DCF produced comparable lateral FWHM estimates to uncoupled DAS. In the phantom, every beamformer produced higher FWHM estimates in elevation than laterally. This was caused by the elevational gap between panels in the Vermon array, essentially representing missing rows of elements. This was especially damaging for NSI, as the required symmetry in elements to form a null was compromised in the elevational direction, causing the elevational FWHM to suffer. If such gaps were not present, NSI should have provided similar resolution improvement in both directions, as demonstrated by the simulations.

Despite resolution benefits, the NSI and DCF beamformers also resulted in trade-offs to sSNR and gCNR, and higher coupling numbers reduced CR. Nevertheless, these results provide evidence that matrix arrays with increased FOV and reduced element count could be produced by increasing the element size. Furthermore, each of these beamformers has methods of tuning the resolution/speckle trade-off, which can help maximize quality for various applications. NSI is tuned by the DC offset \cite{agarwal_improving_2019}, while DCF could be tuned by performing a generalized coherence factor (GCF) calculation on the projections instead of merely CF \cite{li_adaptive_2003}. Speckle quality could be further improved and the contrast enhanced using adaptive tuning methods for NSI \cite{paridar_spatially_2024,yan_contrast-enhanced_2025}. 

The MV beamformer had the best CNR, gCNR, and sSNR metrics behind DAS for each coupling number, but it did not improve resolution for phantom or in vivo experiments. These differences in performance can be explained by considering the fact these beamformers were used in a low-element count scenario. With NSI, the fundamental idea is to form and invert a null \cite{agarwal_improving_2019}. A null can be created using as little as two elements, one positive and one negative, meaning this method should still be effective at improving resolution with fewer elements. However, with MV beamforming, the idea is to calculate apodization weights that minimize the power of the apodized signal while maintaining unit gain at the focal point \cite{synnevag_adaptive_2007}. With fewer elements, and thus fewer apodization weights, the rejection of signals from outside the focal point is reduced, making MV less effective at improving resolution for low-element count scenarios. The DCF beamformer estimates the coherence of signals across directional projections of the matrix elements \cite{wu_directional_2025}. With low element counts, the coherence estimate might be highly noisy, introducing large deviations in the coherence factor between adjacent pixels, and explaining its drastic reduction to sSNR.

Another important consideration for each of these beamformers is the computation that each one requires with respect to the element count. DAS is the most efficient, as it only requires a sum over the elements. In this case, the complexity is O(N) where N is the total number of elements. NSI on a 2D array essentially consists of apodizing delayed channel data six times, one for each apodization (\(ZM_{AZ},ZM_{EL},DC1_{AZ},DC1_{EL},DC2_{AZ},DC2_{EL}\)). The remaining operations, such as the extra envelope detections, the DC averaging (Eq. \ref{eq:avg_dc}), the max over directions (Eqs. \ref{eq:max_zm} and \ref{eq:max_dc}), and subtracting the null (Eq. \ref{eq:subtract_null}) all run in constant time with respect to the elements. Therefore, NSI is also of complexity class O(N), with some added constant factors. The DCF beamformer first requires directional projections (Eq. \ref{eq:projected_vectors}), which are sums over rows or columns of channel data. In other words, it sums \(\sqrt{N}\) values \(\sqrt{N}\) times, making the projection calculation O(N) complexity. The actual coherence factor calculation then only involves summing \(\sqrt{N}\) values from these projected vectors. Therefore, DCF is overall in complexity class O(N). Finally, our MV implementation starts with the same projections, then estimates covariance matrices using spatial smoothing on the projections (Eqs. \ref{eq:projected_covariance_matrices} and \ref{eq:spatially_smoothed_projections}). If the subaperture length for spatial smoothing is denoted by L, then \(L \leq \sqrt{N}/2\), and the size of the covariance matrix is L x L. The most expensive operation in MV is inverting the covariance matrix for the weight calculation (Eq. \ref{eq:mv_weights}). Inverting an L x L matrix is of complexity O(\(L^{3}\)). Therefore, our MV implementation has complexity O(\(L^{3}\)), or O(\(N^{3/2}\)). Based on the complexities, MV will require the most computation, even for low element counts. To examine the differences between the other three beamformers, we include the run times for coupling by two and four with the virtual large aperture in Table \ref{tab5:runtimes}. These represent average run times over 100 iterations on a lab server with an NVIDIA RTX A5000 GPU.

\begin{table}[ht]
    \centering
    \begin{tabular}{|c|c|c|c|c|}
        \hline
        coupling number & DAS & NSI & DCF & MV \\
        \hline
        2 & 30 \(\mu\)s & 23 ms & 15 ms & 1.01 s\\
        4 & 28 \(\mu\)s & 8 ms & 3.8 ms & 438 ms\\
        \hline
    \end{tabular}
    \caption{Run times for different beamformers with element coupling.}
    \label{tab5:runtimes}
\end{table}

\begin{figure}
    \centering
    \includegraphics[width=\linewidth]{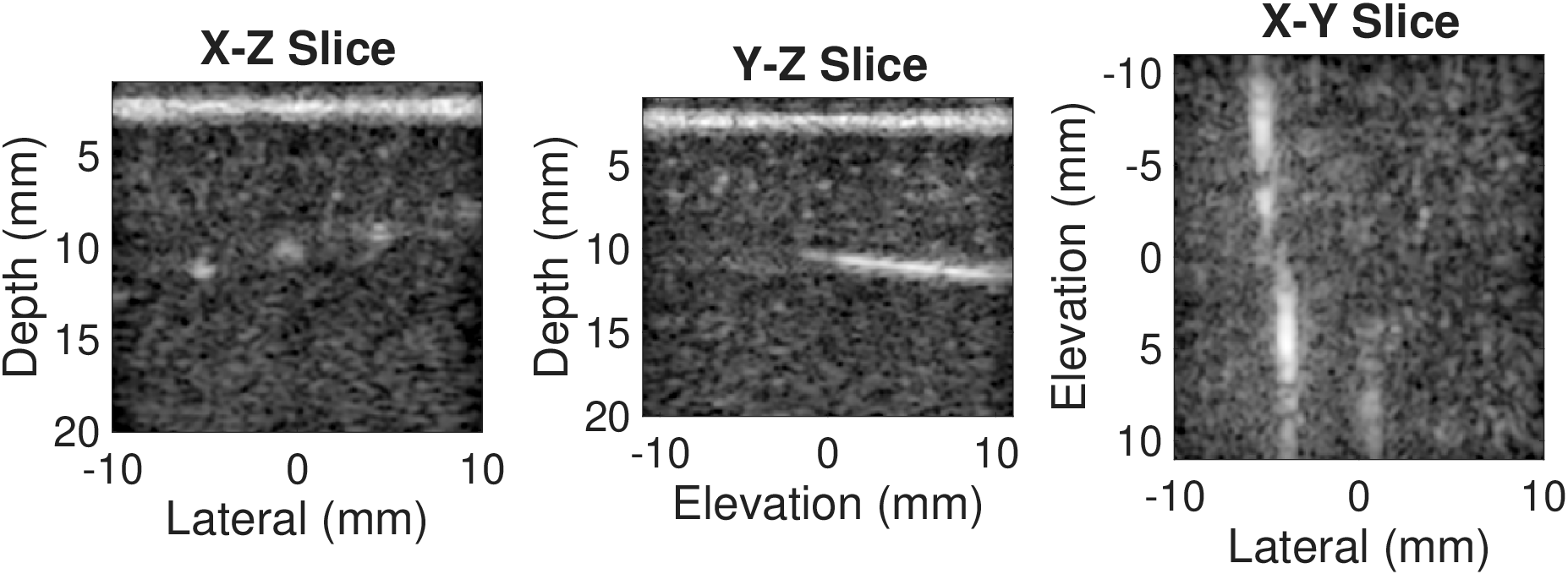}
    \caption{An example of misalignment in the virtual aperture acquisition. The artifacts of misalignment are discontinuities at the boundaries of the four quadrants, and regions only being visible in a portion of orthogonal slices.}
    \label{fig14:misalignement_example}
\end{figure}

Aside from the beamforming, there are also considerations for the limitations of our virtual aperture acquisition and the practicality of the large aperture it imitates. With the virtual aperture acquisition, there is the potential of probe misalignment between quadrants. To illustrate the artifacts of misalignment, we include Figure \ref{fig14:misalignement_example}, which displays B-mode images of wire targets taken on a severely misaligned acquisition. Misalignment in the probe is manifest as discontinuities between acquisition quadrants, and therefore the wire is not fully visible in orthogonal slices aligned with the axes. While our visual alignment cannot guarantee mathematically perfect probe location or rotation, it was enough to keep wires and cysts in entire image slices without the obvious discontinuities displayed in Figure \ref{fig14:misalignement_example}.

As for the practicality of the array size, the virtual aperture we created in this paper has a footprint of 19.2 mm x 21.3 mm. This is comparable to the dimensions of currently available RCA arrays, and with a coupling number of four, it has a comparable element count of 256. Because it's around the the same size as current RCA arrays, it should be fairly straightforward to keep it flat and ergonomically feasible. One disadvantage of RCA arrays is that their extremely long line elements, dozens of wavelengths long, create edge waves that result in ghosting artifacts in plane-wave images \cite{jensen_anatomic_2022}. Our square elements are still small enough, around 6 wavelengths, to not produce such artifacts. Furthermore, line elements can only steer or focus in orthogonal directions, but they cannot steer on diagonals. Square elements can still steer on diagonals (albeit with slightly reduced range), allowing more directional spatial frequency content to be captured by the steered plane-waves. 

Another competitor with our proposed design is recently introduced large square arrays with diverging lenses over each element. Such arrays have primarily been used for Ultrasound Localization Microscopy (ULM). In such arrays, lenses widen the directivity which improves resolution but allows grating lobes to appear due to the large pitch \cite{favre_boosting_2022}. This sacrifice is feasible for ULM, because ULM involves a localization step with can ignore grating lobes, then displays the tracks of localized peaks rather than beamformed echo data \cite{errico_ultrafast_2015}. For any clinical tasks which require visualization of the B-mode data, such as lesion volume estimation, or ROI segmentation (such as for Quantitative Ultrasound), grating lobes cause major artifacts that can obscure the image and disrupt such measurements \cite{paul_side_1997}.

To mention a few potential target applications for our array design, gall bladder volume estimation for detection of biliary diseases \cite{jouleh_comparison_2025}, volume estimation of amniotic fluid \cite{gilja_measurements_1999}, or breast cancer treatment and monitoring \cite{shoma_ultrasound_2006} are all areas that could benefic from increased FOV. Applications where larger apertures are not desirable include endoscopic imaging, where probes need to be small enough to fit inside the body, and echocardiography, which typically employs small footprint phased arrays which can fit between the ribs.

\section{Conclusion}
We have demonstrated how an increased element size, up to 6 \(\lambda\), can allow for twice the FOV in both directions for 2D matrix arrays with reduced element count. The resulting degradation in resolution from increased element size can be mitigated by advanced beamformers such as DCF or NSI, making large elements more viable. However, both these beamformers also reduced speckle and contrast metrics. Further improvement could be made using adaptive tuning methods. The increased FOV we demonstrate can make 3D ultrasound more viable and useful for clinics without increasing the cost or complexity of 3D ultrasound systems.

\section*{Declaration of Competing Interest}
The Authors declare that they have no competing financial interests or personal relationships that appear to have influenced the work reported in this paper.

\section*{Data Availability}
Data will be made available upon request.

\section*{Acknowledgments}
The authors would like to acknowledge and thank Matt Lowerison and YiRang Shin for performing element calibration on the Vermon matrix array.

This work was supported by the National Institutes of Health [grant numbers R21EB035714, R01CA273700, R01EB036800].

\section*{CRediT Authorship Contribution Statement}
\textbf{Mick Gardner:} Writing - original draft; Writing - review and editing; Investigation; Formal Analysis; Software. \textbf{Michael L. Oelze:} Writing - review and editing; Supervision; Project Administration.

\bibliographystyle{elsarticle-num-names} 
\bibliography{large_aperture_2d}

\end{document}